\documentclass[useAMS,usenatbib]{mn2e}
\usepackage{times}

\title[Gravitational lensing by a moderate redshift galaxy group]{The mass distribution of a moderate redshift galaxy group and brightest group galaxy from gravitational lensing and kinematics}
\author[J. P. McKean et al.]{J. P. McKean,$^{1,2}$\thanks{E-mail: mckean@astron.nl} M. W. Auger,$^{3,4}$ L. V. E. Koopmans,$^{5}$ S. Vegetti,$^{5}$ O. Czoske,$^5$
\newauthor C. D. Fassnacht,$^{3}$ T. Treu,$^{4}$ A. More$^{1,6}$ and D. D. Kocevski$^3$\\
$^{1}$Max-Planck-Institut f\"{u}r Radioastronomie, Auf dem H\"{u}gel 69, D-53121 Bonn, Germany\\
$^{2}$ASTRON, Oude Hoogeveensedijk 4, 7991 PD Dwingeloo, the Netherlands\\
$^{3}$Department of Physics, University of California, Davis, CA 95616, USA\\
$^{4}$Department of Physics, University of California, Santa Barbara, CA 93106, USA\\
$^{5}$Kapteyn Astronomical Institute, Postbus 800, NL-9700 AV Groningen, the Netherlands\\
$^{6}$Laboratoire d'Astrophysique de Marseille, UMR 6110, CNRS-Universit{\' e} de Provence, 38 rue Fr{\' e}d{\' e}ric Joliot-Curie, 13388 Marseille Cedex 13, France\\
}
\begin{document}

\date{Accepted 2010 January 11. Received 2010 January 11; in original form 2009 October 6}

\pagerange{\pageref{firstpage}--\pageref{lastpage}} \pubyear{2008}

\maketitle

\label{firstpage}

\begin{abstract}
The gravitational lens system CLASS B2108+213 has two radio-loud lensed images separated by 4.56 arcsec. The relatively large image separation implies that the lensing is caused by a group of galaxies. In this paper, new optical imaging and spectroscopic data for the lensing galaxies of B2108+213 and the surrounding field galaxies are presented. These data are used to investigate the mass and composition of the lensing structure. The redshift and stellar velocity dispersion of the main lensing galaxy (G1) are found to be $z=$~0.3648\,$\pm$\,0.0002 and $\sigma_v =$~325\,$\pm$\,25~km\,s$^{-1}$, respectively. The optical spectrum of the lensed quasar shows no obvious emission or absorption features and is consistent with a BL Lac type radio source. However, the tentative detection of the G-band and Mg-b absorption lines, and a break in the spectrum of the host galaxy of the lensed quasar gives a likely source redshift of $z=$~0.67. Spectroscopy of the field around B2108+213 finds 51 galaxies at a similar redshift to G1, thus confirming that there is a much larger structure at $z\sim$~0.365 associated with this system. The width of the group velocity distribution is 694\,$\pm$\,93~km\,s$^{-1}$, but is non-Gaussian, implying that the structure is not yet viralized. The main lensing galaxy is also the brightest group member and has a surface brightness profile consistent with a typical cD galaxy. A lensing and dynamics analysis of the mass distribution, which also includes the newly found group members, finds that the logarithmic slope of the mass density profile is on average isothermal inside the Einstein radius, but steeper {\it at} the location of the Einstein radius. This apparent change in slope can be accounted for if an external convergence gradient, representing the underlying parent halo of the galaxy group, is included in the mass model.
\end{abstract}


\section{Introduction}
 
In the standard model of galaxy formation, structure arises hierarchically through the mergers and interactions of smaller mass systems, leading to increasingly more massive structures. Cold dark matter (CDM) simulations of these merging events predict universal radial matter density profiles with inner slopes of $\gamma \sim$~1--1.5 [where $\rho_{\rm CDM}(r) \propto r^{-\gamma}$], which are independent of the halo mass [e.g. \citealt*{navarro96} (NFW);
\citealt{moore98}; \citealt*{diemand07,navarro09}]. Furthermore, the compact cores of merging systems are expected to survive the cannibalism process, forming CDM substructure within the new parent halo (e.g. \citealt{moore99,gao04,diemand07}).

Strong gravitational lensing, where multiple images of the same background object are formed by an intervening potential, has provided a powerful observational test of these theoretical predictions for mass structures out to $z \sim$~1, particularly when combined with stellar kinematic data (e.g. \citealt{treu04,sand04}; \citealt{koopmans06}). 

In the case of galaxy-scale lensing ($M_{\rm lens}=$\,10$^{11}$--10$^{12}~$M$_{\odot}$), detailed mass modeling of the positions and flux densities of the multiply imaged source has shown the total (i.e. baryons plus dark matter; $\rho_{\rm tot}$) mass density profiles of lens galaxies to be close to isothermal, that is, $\gamma'=d \log \rho_{\rm tot}/d\log r \sim$~2 (\citealt{cohn01}; \citealt{koopmans06}; \citealt*{rusin03a,treu04,wucknitz04}; \citealt{koopmans09}) with very little scatter ($\sim$5 per cent on $\gamma'$).  The steeper slope than that predicted by dark matter only simulations is generally attributed to baryon cooling and collapse, which steepens the inner density profile for lens galaxies \citep{blumenthal84,gnedin04,keeton01b,kochanek01a,jiang07}. The origin of the small scatter in the total mass density profile slope is not completely clear (see discussion in \citealt{koopmans06}).

Decomposing the baryonic and dark matter components is particularly hard for galaxy-scale lenses, given the dominance of baryonic mass at small radii. By combining lensing with stellar kinematic tracers, and assuming a spatially uniform mass-to-light ratio for the stellar distribution, an upper limit of $\gamma \sim$~1.3 for the dark matter component of galaxy-scale haloes \citep{treu04} can be obtained, which is certainly within the allowable range predicted from simulations. However, studies of the inner slopes of dark matter haloes using non-lensing techniques, for example, from rotation curve measurements of nearby dwarf and spiral galaxies, find evidence for a flattened core (0 $< \gamma <$ 1) out as far as the optical radius (e.g. \citealt{salucci00,swaters03,gentile04,donato09}), which are typically inconsistent with the CDM models. Although, it should be noted that due to systematic effects, such as non-circular motions or poorly defined disk inclinations, the CDM models cannot be discounted \citep{swaters03}.

For galaxy clusters ($M_{\rm lens}>$\,10$^{15}$~M$_{\odot}$), the situation is significantly different. On the one hand, the stellar component is less important than for galaxies on the scales that are relevant to measure the inner slope of the dark matter, and several mass probes are available to cover the broad range of radii needed for this measurement, including strong lensing, stellar kinematics, X-rays and weak lensing. On the other hand, many clusters are not dynamically relaxed systems, often with significant triaxiality and substructure. Therefore the interpretation of the results can be more complicated than in the case of galaxies and can be affected by potential systematics, such as those arising from projection effects and departures from hydrostatic equilibrium.

A relatively large scatter of inner density profile slopes has been reported for galaxy clusters (\citealt{smith01}; \citealt*{sand02}; \citealt*{ettori02}; \citealt*{lewis03}; \citealt{sand04,sand05}; \citealt{gavazzi05}; \citealt*{saha06}; \citealt{sharon05}; \citealt{bradac08}) -- ranging from flat inner cores to slopes consistent with CDM predictions. However, it should be noted that some of these authors refer to the slope of the dark matter component ($\gamma$) or to the slope of the total mass density profile ($\gamma'$) as if they were the same quantity. This is not justified. In fact, although stars are only a small fraction of the total mass, they are extremely spatially concentrated and therefore they affect the {\it slope} in a significant way (\citealt{sand02}). Overall, although more work remains to be done at cluster scales, it seems clear that in at least some cases the inner slope of the dark matter halo is flatter than predicted by CDM simulations and the NFW profile is not a good fit to the data (\citealt{sand08}).

Galaxy groups ($M_{\rm lens}=$\,$10^{13}$--$10^{14}~$M$_{\odot}$), containing a few to a few tens of galaxies, are the most common environment for galaxies to reside at low redshifts (e.g. \citealt{eke04}). Although the total mass of a galaxy group can be estimated from the luminosity distribution or the stellar mass of the individual group members (e.g. \citealt{yang07}) as well as from the group velocity dispersion \citep{zabludoff98}, the distribution of the matter within groups is not well known. 

Galaxy groups have been relatively unexplored with strong gravitational lensing due to the paucity of known direct image splitting by group sized haloes (see \citealt{limousin09} and \citealt{kubo09} for recent developments) and weak lensing measurements can require the stacking of possibly hundreds to thousands of groups (e.g. \citealt{hoekstra01,parker05,mandelbaum06}), and even then, only probe the outer part of the dark matter halo. There are several examples where strong gravitational lensing galaxies have been found to be associated with a group of galaxies and/or there are groups of galaxies found along the line-of-sight to the lensed object \citep{tonry98,tonry99,tonry00,fassnacht02,faure04,morgan05,fassnacht06,momcheva06,williams06,auger07}. In these cases, the lensed image separations are still in the galaxy-scale regime of $\sim$\,0.5--1.5 arcsec because the lensing potential is dominated by a single galaxy, with the contribution of the group to the overall convergence being only $\sim$5 per cent \citep{momcheva06,auger07,auger08}. However, it should be noted that environmental contributions at this level can still have a large impact on the results of lens modelling \citep{keeton04}. For lens systems with image separations larger than 3~arcsec the contribution of the environment is thought to be much larger than in the galaxy-scale regime (with an external convergence $\ga$~10 per cent), possibly enhancing the lensing probabilities \citep*{oguri05}. 

How the mass is distributed within a group-sized halo, and what effect baryons have on the overall density profile is not entirely clear. However, as the image separation increases there is expected to be a transition from galaxy scale haloes, which are dominated by stars in the centre and have a relatively steep total mass density profile (i.e. $\gamma' \sim$~2), to group-scale haloes, characterized by a flatter inner density profile of $\gamma' \sim$~1--1.5 \citep{oguri06}. Another question to be addressed is whether the dense environment affects the dark matter sub-haloes of the satellite galaxies within the group. It is expected that due to tidal interactions, the dark matter of the sub-haloes will be stripped, resulting in a change in the sub-halo mass-to-light ratio (e.g. \citealt{gao04}) and a truncation which would lead to a more concentrated sub-halo mass distribution (e.g. \citealt{bullock01}).

In this paper, we present for the first time an analysis of the total (baryonic and dark) inner density profile of a moderate redshift galaxy group using strong gravitational lensing and stellar kinematics. We show that the gravitational lens system B2108+213 is part of a massive group of galaxies and that the multiple imaging is found around the large central galaxy, as opposed to the satellite members of the group. Thus, we are probing the inner part of the group-scale halo. In Section \ref{classb2108+213}, we present an introduction to the lens system and summarize the previous observations. In Section \ref{photometry}, new ground based imaging with the W. M. Keck Telescope is combined with archival {\it HST} data to select potential members of the galaxy group associated with B2108+213. Our spectroscopy of group candidates is presented in Section \ref{spectroscopy}. From these observations we found 51 galaxies at the same redshift as the main lensing galaxy. The properties of the group and the constituent group members are discussed in Section \ref{redshifts}. An analysis of the stellar velocity dispersion and surface brightness profile of the main lensing galaxy is presented in Section \ref{kinematics}. In Section \ref{modelling} we present a new mass model for B2108+213 which incorporates the group galaxies and the stellar kinematics. We discuss our results in Section \ref{disc} and present our conclusions in Section \ref{conc}.

For all calculations we adopt an $\Omega_{M} =$~0.3, $\Omega_{\Lambda}=$~0.7 spatially-flat Universe, with a Hubble constant of $H_{0} =$~70~km\,s$^{-1}$~Mpc$^{-1}$. At redshift 0.365, 1 arcsec corresponds to an angular size of 5.075~kpc.

\section{CLASS B2108+213}
\label{classb2108+213}

The gravitational lens system CLASS B2108+213 \citep{mckean05}, which was discovered in the final phase of the Cosmic Lens All-Sky Survey (CLASS; \citealt{myers03}; \citealt{browne03}), consists of at least two lensed images of a radio-loud quasar that are separated by 4.56~arcsec. The wide image separation between the two lensed images implies lensing by a massive halo, due to a group-sized mass distribution. Therefore, B2108+213 provides an excellent opportunity to probe a mass regime where very little is known. The flux-ratio of the two lensed images (A and B; where image A has the higher flux-density) is $\sim$2:1. High resolution radio imaging at 1.7 and 5~GHz with Very Long Baseline Interferometry (VLBI) found a compact core and extended jet emission in both lensed images, which is consistent with gravitational lensing \citep{more07}. From optical and infrared imaging with the {\it Hubble Space Telescope} ({\it HST}) a massive early-type elliptical galaxy (G1) was found between the two lensed images, which is thought to dominate the lensing potential. However, there is an additional satellite galaxy (G2) within the Einstein radius and a scatter of nearby galaxies of similar colour to G1 within $\sim$15~arcsec of the system. In addition, the infrared emission from lensed image A shows a large gravitational arc which is due to the lensed emission from the host galaxy of the background quasar \citep{mckean05}. Observations with {\it Chandra} revealed luminous X-ray emission, which is presumably from the hot group/cluster gas, that is elongated by $\sim$~1~Mpc in an east--west direction across the lens system \citep{fassnacht08}. There is also a third radio component (C) which is coincident with the optical emission from G1 to within the astrometric uncertainties. However, component C is not thought to be a third lensed image because, first, further radio imaging with VLBI and the Multi-Element Radio Linked Interferometry Network (MERLIN) found compact and extended radio emission consistent with an Active Galactic Nucleus (AGN) embedded within the main lensing galaxy G1, and second, the flux-density of component C is more than 10$^6$ times too high when compared to what is expected from modelling the lensing mass distribution \citep{more07}.

A simple lens model, which includes both G1 and G2 with isothermal density profiles and an external shear, can explain the positions of the lensed images A and B, but fails to reproduce the observed flux-ratio \citep{mckean05,more07}. This could be explained if there is a perturbation in the lensing mass model, for example, due to mass substructure in the parent dark matter halo (e.g. \citealt{bradac02,dalal02,metcalf02,vegetti09b}) or from the contribution of the nearby group galaxies. The best fitting model for the system that satisfies the positional and flux constraints requires the combined luminous and dark matter density profile of G1 to be steeper than isothermal ($\gamma' =$~2.45$^{+0.19}_{-0.18}$ at the 68 per cent confidence level; \citealt{more07}).

\begin{table}
\begin{center}
\caption{Observing logs for the photometric and spectroscopic (long- and multi-slit) observations of the B2108+213 field. These observations were carried out with the W. M. Keck-I Telescope. Slit position angles (PA) are given east of north.}
\begin{tabular}{llccr} \hline
Filter or slit/       & \multicolumn{1}{c}{Date}        & \multicolumn{2}{c}{Exposure time ($s$)} & \multicolumn{1}{c}{PA}  \\ 
mask number   &				& LRIS-B    	    	& LRIS-R		& \multicolumn{1}{c}{($\degr$)} \\ \hline
{\it U}		& 2003 Oct 24/25	& 3400			&			&  334 \\
{\it B}			& 2003 Oct 24/25	& 2400       		&			&  334 \\
{\it V}			& 2003 Oct 24/25	&            			& 600		&  334 \\
{\it R}		& 2003 Oct 24/25	&            			& 2000		&  334 \\
{\it I}			& 2003 Oct 24/25	&            			& 2400		&  334 \\
Slit 1        	 	& 2003 Oct 24		& 5400          		& 5400		&  142 \\
Slit 2          	& 2003 Oct 25		& 2500         	 	& 2500		&  60  \\
Slit 3          	& 2003 Oct 25		& 2400         	 	& 2400		&  25  \\
Slit 4          	& 2003 Oct 25		& 1800         	 	& 1800		&  75  \\
Slit 5			& 2007 Oct 10		& 	     			& 2400		&  50  \\
Mask 1		& 2004 Jun 15		& 3600			& 3600		&  334 \\
Mask 2		& 2004 Aug 12		& 5400			& 5400		&  334 \\
Mask 4		& 2004 Aug 12		& 3600			& 3600		&  334 \\
Mask 5		& 2005 Aug 1		& 5400			& 5400		&  345 \\
Mask 6		& 2005 Aug 1		& 5400			& 5400		&  345 \\
Mask 7		& 2005 Aug 1		& 5400			& 5400		&  345 \\
\hline
\end{tabular}
\label{log}
\end{center}
\end{table}

\section{Photometry}
\label{photometry}

The first step of our investigation into the local environment and lensing galaxies of B2108+213 was to obtain deep imaging of the field using the W. M. Keck Telescope. These observations were designed to identify galaxies which could potentially be part of a group or cluster associated with the main lensing galaxy, G1.

\subsection{W. M. Keck Telescope imaging}
\label{photo-keck}

We carried out extensive optical imaging of the B2108+213 field with the W. M. Keck-I Telescope on 2003 October 24 and 25. The conditions on both nights were not photometric due to faint cirus cloud cover. Therefore, photometric redshifts of the field galaxies could not be calculated. However, the relative colours could still be used to select group candidates for follow-up spectroscopy. The full width at half maximum (FWHM) seeing was $\sim$0.7~arcsec throughout the observing run. Our imaging data were taken with the Low Resolution Imaging Spectrograph (LRIS; \citealt{oke95}), which allows simultaneous observations through blue (-B) and red (-R) optimized cameras. A 5600~{\AA} dichroic was used to split the light between LRIS-B and -R for our {\it U}, {\it B}, {\it R} and {\it I}-band filter photometry. For our {\it V}-band imaging, we employed a 4600~{\AA} dichroic. With these filters in place, the usable field of view through LRIS-B and -R is $\sim$\,5.5\,$\times$7.5~arcmin$^{2}$. The integration times through each filter were typically $n$\,$\times$\,300~s, with a 10 arcsec dither between each exposure to aid the removal of bad pixels and cosmic rays during the reduction stage. A summary of our imaging observations is given in Table \ref{log}.

The data were reduced within {\sc iraf}\footnote{{\sc iraf} ({\sc i}mage {\sc r}eduction and {\sc a}nalysis {\sc f}acility) is distributed by the National Optical Astronomy Observatories, which are operated by AURA, Inc., under cooperative agreement with the National Science Foundation.} using standard procedures. The world coordinate system was added to each image by fitting to the known positions of United States Naval Observatory (USNO) stars. This resulted in an astrometric uncertainty of $\sim$0.3~arcsec. The photometric catalogue was generated by {\sc sextractor} \citep*{bertin96}.

No zero-point magnitude calibration or atmospheric extinction correction was applied to the data because the observing conditions were not photometric. However, we estimated the apparent magnitudes of the objects in the field by bootstrapping an approximate calibration to the {\it B} and {\it R} band imaging using unsaturated stars which were detected in the USNO B1 catalogue. We used archival {\it HST} imaging at F555W and F814W (see Section \ref{hst}) to apply a photometric calibration to the {\it V} and {\it I} band datasets.

\subsection{Hubble Space Telescope imaging}
\label{hst}

{\it HST} imaging of the B2108+213 field with the Advanced Camera for Surveys (ACS) and the Near Infrared Camera/Multi-Object Spectrograph (NICMOS) was presented by \citet{mckean05}. To summarise, B2108+213 was observed with the ACS through the F555W and F814W filters. Both observations were carried out in the wide field channel mode, which provided a field of view of 202\,$\times$\,202~arcsec$^{2}$. An infrared image at F160W was aquired with NICMOS using the NIC2 camera. However, infrared information is only available for 12 sources nearby to B2108+213 because the NIC2 field of view is 19\,$\times$\,19~arcsec$^{2}$. The photometric analysis of the ACS and NICMOS data was carried out with {\sc sextractor}.

\section{Spectroscopy}
\label{spectroscopy}

Spectroscopic observations of B2108+213 and the candidate group galaxies were carried out with LRIS on the W. M. Keck-I Telescope in both long- and multi-slit modes.
 
\subsection{Long slit spectroscopy}

\begin{figure}
\begin{center}
\setlength{\unitlength}{1cm}
\begin{picture}(6,17.2)
\put(-1.5,-0.3){\includegraphics{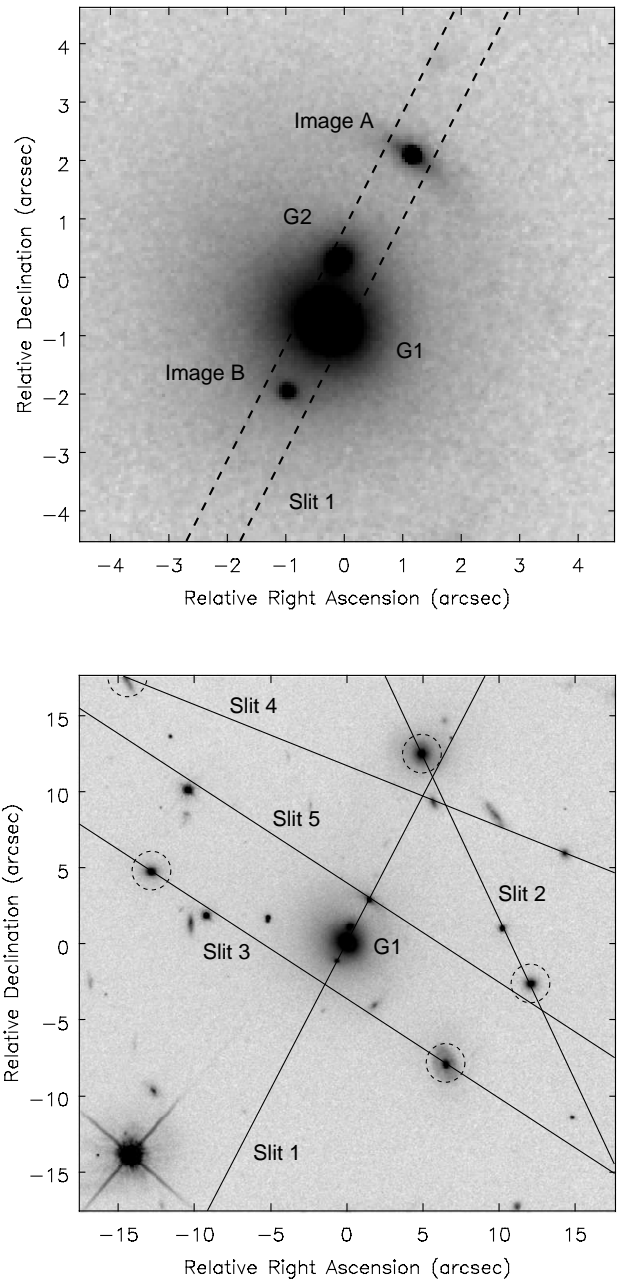}}
\end{picture}
\caption{({\it Top}) The gravitational lens system B2108+213 and the position of Slit 1. ({\it Bottom}) The slit positions for the long-slit spectroscopy carried out across the field of B2108+213 (underlayed is the {\it HST} ACS-F814W image). Those galaxies in dashed circles are part of the group associated with the main lensing galaxy (G1) at a redshift of 0.365. The area of sky shown is 180$\times$180~kpc$^2$ at the redshift of G1.}
\label{slit-pos}
\end{center}
\end{figure}

The long slit data were taken with LRIS during two observing sessions. The first set of spectroscopic observations were carried out at the same time as the LRIS photometry presented in Section \ref{photo-keck} (2003 October 24 and 25). In total, we employed four 1 arcsec wide and 120 arcsec long slits across the field (see Fig. \ref{slit-pos} for the slit positions). Slit 1 was positioned and orientated to include the B2108+213 lensing galaxies and lensed images. The three remaining long slits (Slits 2 to 4) targeted those galaxies with similar colours to G1 (based on an initial analysis of the LRIS imaging data done at the telescope) and in close proximity to B2108+213. The second set of spectroscopic observations were carried out duing 2007 October 10. Here, a single 1 arcsec-wide long slit (slit 5; see Fig. \ref{slit-pos}) was placed across the gravitational arc of lensed image A. The aim of this observation was to measure the redshift of the background source. A summary of the total exposure times and position angles used for each slit is presented in Table \ref{log}. 

For the observations taken on 2003 October 24 and 25, the light was split between LRIS-B and -R using a 5600~{\AA} dichroic. A 600~lines~mm$^{-1}$ grism blazed to 4000~{\AA} was used through LRIS-B. This produced a dispersion of 0.61~{\AA}~pixel$^{-1}$ and a wavelength coverage of $\sim$3400 to 5600~{\AA}. The light was dispersed through LRIS-R using a 600~grooves~mm$^{-1}$ grating centred at 7050~{\AA}. This resulted in an LRIS-R dispersion of 1.26~{\AA}~pixel$^{-1}$ and a wavelength range of $\sim$5660 to 8250~{\AA}. The wavelength calibration was established using Hg, Cd, Ne and Ar arc exposures. An observation of the spectrophotometric standard star BD+28$\degr$4211 \citep{oke90} was taken on each night to correct for the response of the instrument. For the observations taken on 2007 October 10, the 5600~{\AA} dichroic was used with the 400~grooves~mm$^{-1}$ grating centred at 8500~{\AA} through LRIS-R. This gave a dispersion of 1.86~{\AA}~pixel$^{-1}$ and a wavelength range of $\sim$5700 to 9200~{\AA}. Note that no standard star observation was carried out so the resulting spectrum of lensed image A was not flux-calibrated. The one-dimensional spectrum of each detected object was reduced within {\sc iraf} following the procedures described in \citet{mckean04}.

Our long slit observations yielded spectra for 16 galaxies and 2 stars. Eight galaxies (including G1) were found to be at a redshift of $\sim$0.365; thus confirming that the main lensing galaxy of B2108+213 is part of a larger structure. A detailed discussion of the results from our long slit spectra is given in Section \ref{redshifts}.

\subsection{Multiple slit spectroscopy}

Our long slit observations only sampled a small number of galaxies within $\sim$0.3~Mpc of B2108+213. However, the full extent of the group or cluster associated with the B2108+213 lensing potential may be much larger. Therefore, we undertook multiple slit spectroscopy of the field using LRIS on 2004 June 15, 2004 August 12 and 2005 August 1. 

\begin{figure}
\begin{center}
\setlength{\unitlength}{1cm}
\begin{picture}(6,8.0)
\put(-1.2,-2.2){\includegraphics{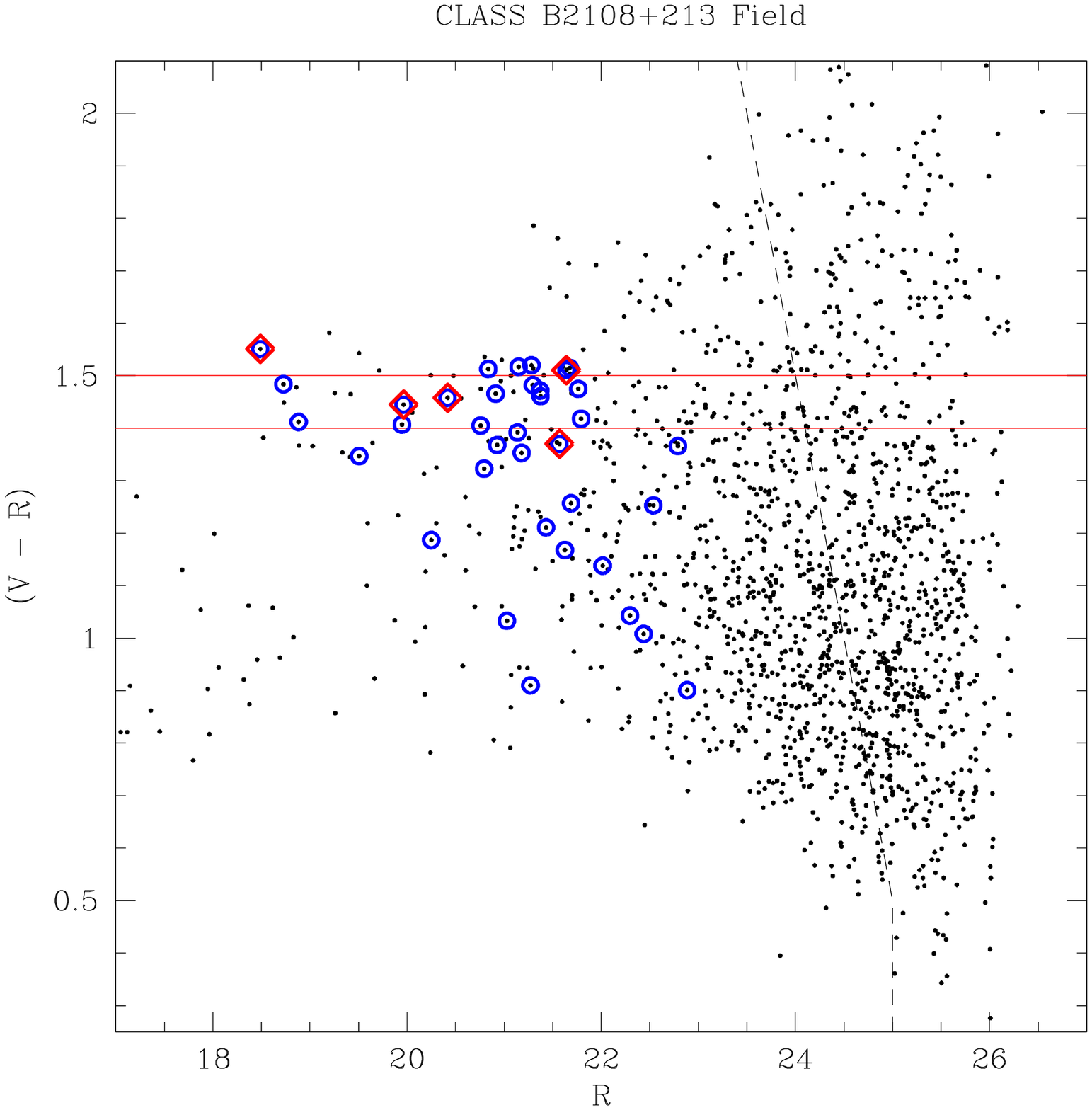}}
\end{picture}
\caption{The {\it V}$-${\it R} colour {\it R} band magnitude plot of the B2108+213 field. The red diamonds mark the observed colours of G1 and the 4 closest galaxies to G1 detected from the long-slit observations, whereas the blue circles mark those objects with measured redshifts. Note that this relative colour plot is made from non-photometric data and the expected {\it V}$-${\it R} colour of an elliptical galaxy at redshift 0.365 is 1.2.}
\label{cmVR}
\end{center}
\end{figure}

\begin{figure}
\begin{center}
\setlength{\unitlength}{1cm}
\begin{picture}(6,8.0)
\put(-1.2,-2.2){\includegraphics{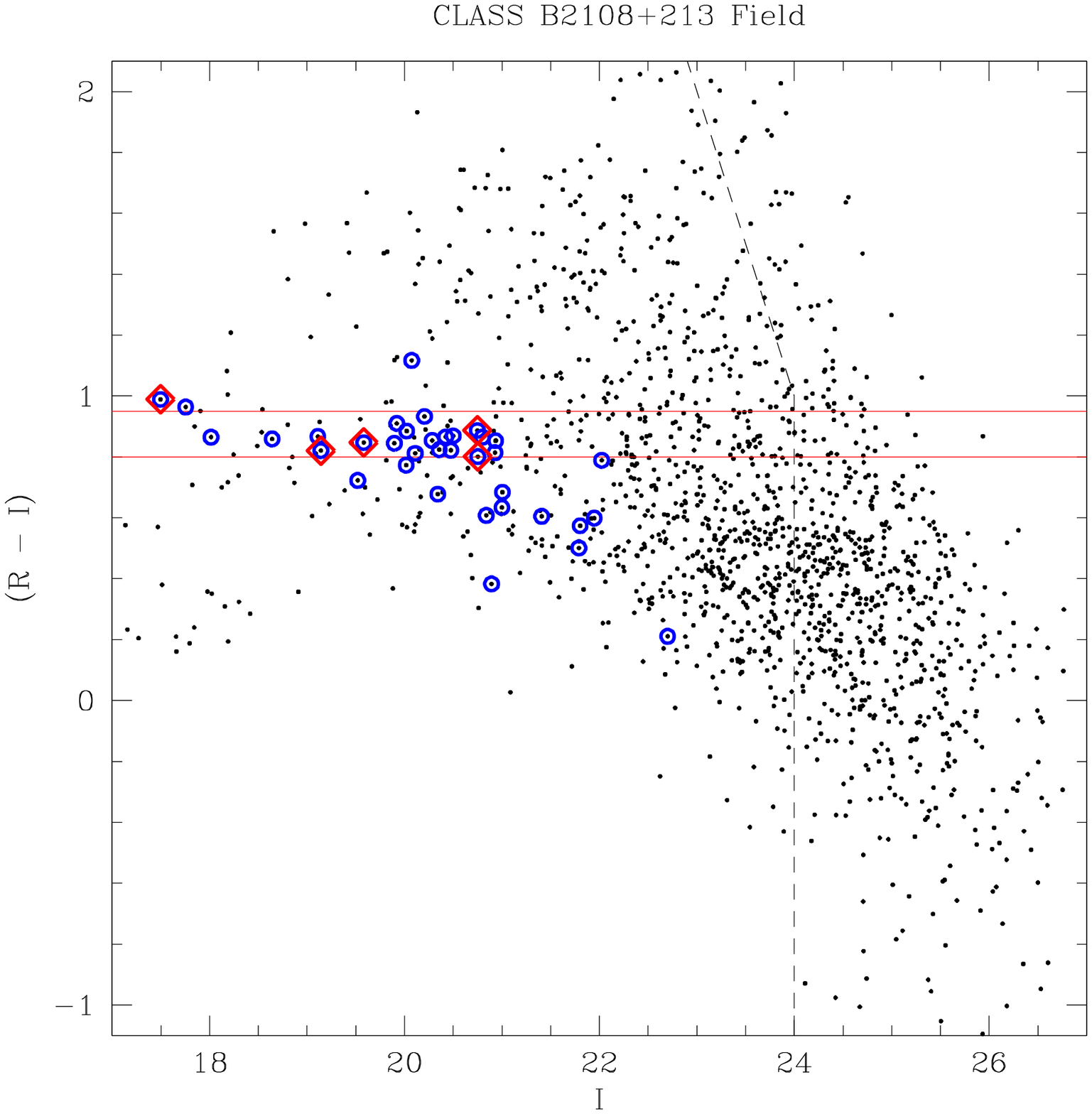}}
\end{picture}
\caption{The {\it R}$-${\it I} colour {\it I} band magnitude plot of the B2108+213 field. The red diamonds mark the observed colours of G1 and the 4 closest galaxies to G1 detected from the long-slit data, whereas the blue circles mark those objects with measured redshifts. Note that this relative colour plot is made from non-photometric data and the expected {\it R}$-${\it I} colour of an elliptical galaxy at redshift 0.365 is 0.9. }
\label{cmRI}
\end{center}
\end{figure}

The multiple slit capability of LRIS allows the spectra of $\sim$25--30 objects to be obtained simultaneously within a 8\,$\times$\,6 arcmin$^{2}$ field through the use of slit masks. Candidate group galaxies were selected primarily by colour. In Figs. \ref{cmVR} and \ref{cmRI} we show the {\it V}$-${\it R} and {\it R}$-${\it I} colour--magnitude diagrams with respect to {\it R} and {\it I} for the B2108+213 field. We stress that these LRIS imaging data were taken on non-photometric conditions and as such the absolute colours provide limited information about a galaxy's redshift. However, it is still possible to use the relative colours to identify galaxies at similar redshifts. The 5 closest  group members to the lensed images discovered from our long-slit observations have been labelled with red diamonds in Figs. \ref{cmVR} and \ref{cmRI}; there is clear evidence of a red sequence in both color--magnitude diagrams. On Figs. \ref{cmVR} and \ref{cmRI} we have also defined the {\it V}$-${\it R} and {\it R}$-${\it I} colour boundaries which were used to select candidate group galaxies. These colour boundaries were set to include all of the group galaxies found from the long-slit data, with the exception of the bluest and reddest galaxies. Targets were chosen for the slit-masks using four priority classes; that is, galaxies meeting i) both the {\it V}$-${\it R} and {\it R}$-${\it I} colour ranges, ii) only the {\it V}$-${\it R} colour range, iii) only the {\it R}$-${\it I} colour range, and iv) neither the {\it V}$-${\it R} and {\it R}$-${\it I} colour ranges. The final class was used to ensure as many spectra as possible were taken on each slit mask.

The first set of multiple-slit data were taken during poor observing conditions on 2004 June 15. The 400 lines~mm$^{-1}$ grism blazed to 3400~{\AA} (1.09~\AA~pixel$^{-1}$ dispersion) and the 400 lines~mm$^{-1}$ grating (1.86~\AA~pixel$^{-1}$ dispersion) centred to 6800~{\AA} was used through LRIS-B and -R, respectively. The light was split with the 4600~{\AA} dichroic. Spectroscopy with a single slit mask (mask 1) for 1~h targeted 32 galaxies in the B2108+213 field. The second set of observations were carried out on 2004 August 12, again during poor observing conditions. The 600 lines~mm$^{-1}$ grism blazed to 4000~{\AA} (0.63~\AA~pixel$^{-1}$ dispersion) was used with LRIS-B, and the 600 lines~mm$^{-1}$ grating (1.28~\AA~pixel$^{-1}$ dispersion) centred to 7000~{\AA} was used through LRIS-R. Two slit masks (mask 2 and mask 4) targeted 60 galaxies. The total observing time for each mask was 1.5 and 1~h, respectively. The third set of multiple-slit observations were carried out during clear conditions on 2005 August 1. Here, three slit-masks were employed, which targeted 77 galaxies in the B2108+213 field. We also placed a single slit over both G2 and lensed image A in each of the three slit masks (total exposure time of 4.5 h) in an attempt to measure their redshifts. The 300 lines~mm$^{-1}$ grism blazed to 5000~{\AA} (1.43~\AA~pixel$^{-1}$ dispersion) and the 400 lines~mm$^{-1}$ grating\footnote{The central wavelength was 7974~{\AA} for Mask 5, and 8042~{\AA} for Masks 6 and 7.} (1.86 \AA~pixel$^{-1}$ dispersion) were used with LRIS-B and R, respectively. The 5600~{\AA} dichroic was used to split the light. A summary of the multiple slit observations is given in Table~\ref{log}. 

The LRIS slit masks were reduced using a custom pipeline (see \citealt{auger08} for details) that bias subtracts, flat-fields, and wavelength-calibrates the spectra before extracting and combining the one-dimensional spectra for each slit. Redshifts were determined for the extracted spectra by employing an automated cross-correlation with galaxy template spectra and manually confirming or correcting each of these redshifts by eye. In total, 154 independent objects were observed with the slit masks (note that some of these objects were also observed previously with long slits). Redshifts were obtained for 84 extragalactic objects and 21 stars were identified.

\section{Redshifts}
\label{redshifts}

The combined long and multiple-slit datasets yielded redshifts for 90 galaxies. The positions, redshifts and {\it HST} magnitudes (F555W and F814W) of each galaxy are given in Table \ref{spec-results}. We now give a description of the spectra for the two lensing galaxies within the Einstein radius of the system and the lensed source. We also discuss the properties of the large-scale structure associated with G1.

\begin{table*}
\begin{center}
\caption{Positions, redshifts and spectral properties of galaxies with confirmed redshifts.}
\label{spec-results}
\begin{tabular}{cccccccc} \hline
Object               & RA             & Dec           & Redshift & Mask     & Emission line(s)									& F555W 		& F814W \\ 
		        & (J2000)	  & (J2000)	      &                 &               &                                                                                                                &                          &               \\ \hline
J211042.75+213246.3  &  21 10 42.750  &  21 32 46.27  &  0.7033  &  2108m2  &  absorption lines only									&			&\\
J211045.30+213204.3  &  21 10 45.304  &  21 32 04.32  &  0.2875  &  2108m7  &  [O\,{\sc ii}],H$\beta$,[O\,{\sc iii}] 						& 20.91\,$\pm$\,0.03	& 19.54\,$\pm$\,0.03\\
J211046.38+213403.5  &  21 10 46.376  &  21 34 03.47  &  0.3639  &  2108m7  &  [O\,{\sc ii}] 									&			&\\
J211047.79+212856.4  &  21 10 47.787  &  21 28 56.44  &  0.3930  &  2108m7  &  [O\,{\sc ii}],H$\beta$ 								&			&\\
J211047.79+213345.8  &  21 10 47.789  &  21 33 45.78  &  0.3628  &  2108m7  &  [O\,{\sc ii}],H$\beta$,[O\,{\sc iii}]						&			&\\
J211047.80+213137.2  &  21 10 47.797  &  21 31 37.23  &  0.5281  &  2108m7  &  [O\,{\sc ii}] 									& 23.61\,$\pm$\,0.04	& 21.10\,$\pm$\,0.03\\
J211048.97+213151.9  &  21 10 48.973  &  21 31 51.89  &  0.7028  &  2108m2  &  [O\,{\sc ii}] 									& 24.09\,$\pm$\,0.04	& 21.31\,$\pm$\,0.03\\
J211049.73+212803.2  &  21 10 49.729  &  21 28 03.17  &  0.5295  &  2108m7  &  absorption lines only 									&			&\\
J211049.80+213341.8  &  21 10 49.795  &  21 33 41.76  &  0.1405  &  2108m7  &  [O\,{\sc ii}],H$\beta$,[O\,{\sc iii}],H$\alpha$,[N\,{\sc ii}],[S\,{\sc ii}] 	&			&\\
J211049.84+213053.9  &  21 10 49.839  &  21 30 53.85  &  0.2727  &  2108m7  &  [O\,{\sc ii}],H$\gamma$,H$\beta$,[O\,{\sc iii}],H$\alpha$,[N\,{\sc ii}] 		& 21.97\,$\pm$\,0.03	& 20.38\,$\pm$\,0.03\\
J211049.99+213318.5  &  21 10 49.992  &  21 33 18.53  &  0.6539  &  2108m7  &  [O\,{\sc ii}],H$\beta$,[O\,{\sc iii}] 						& 23.94\,$\pm$\,0.04	& 21.81\,$\pm$\,0.03\\
J211050.31+213110.1  &  21 10 50.314  &  21 31 10.11  &  0.3631  &  2108m7  &  absorption lines only 									& 24.06\,$\pm$\,0.04	& 22.14\,$\pm$\,0.03\\
J211050.34+213025.6  &  21 10 50.341  &  21 30 25.61  &  0.5607  &  2108m5  &  [O\,{\sc ii}] 									& 23.80\,$\pm$\,0.04	& 21.31\,$\pm$\,0.03\\
J211050.69+213420.6  &  21 10 50.693  &  21 34 20.65  &  0.3630  &  2108m5  &  [O\,{\sc ii}],H$\beta$,[O\,{\sc iii}] 						&			&\\
J211050.86+212948.6  &  21 10 50.856  &  21 29 48.62  &  0.3667  &  2108m5  &  absorption lines only 									& 23.80\,$\pm$\,0.04	& 21.11\,$\pm$\,0.03\\
J211051.32+212949.1  &  21 10 51.321  &  21 29 49.12  &  0.3646  &  2108m7  &  absorption lines only 									& 23.97\,$\pm$\,0.04	& 21.99\,$\pm$\,0.03\\
J211051.58+213003.6  &  21 10 51.582  &  21 30 03.58  &  0.3668  &  LSlit   &  absorption lines only 									& 23.37\,$\pm$\,0.04	& 21.13\,$\pm$\,0.03\\
J211051.77+212929.1  &  21 10 51.768  &  21 29 29.14  &  0.6971  &  2108m7  &  [O\,{\sc ii}],H$\beta$ 								&			&\\
J211051.83+212936.9  &  21 10 51.826  &  21 29 36.94  &  0.3656  &  2108m5  &  [O\,{\sc ii}] 									& 22.42\,$\pm$\,0.03	& 20.29\,$\pm$\,0.03\\
J211052.02+213358.2  &  21 10 52.017  &  21 33 58.24  &  0.3652  &  2108m2  &  absorption lines only									&			&\\
J211052.44+213236.6  &  21 10 52.438  &  21 32 36.57  &  0.3674  &  2108m6  &  absorption lines only 									& 22.49\,$\pm$\,0.03	& 20.16\,$\pm$\,0.03\\
J211052.52+212722.4  &  21 10 52.515  &  21 27 22.36  &  0.8197  &  2108m7  &  [O\,{\sc ii}] 									&			&\\
J211052.86+212851.2  &  21 10 52.861  &  21 28 51.25  &  0.3641  &  2108m5  &  absorption lines only									&			&\\
J211053.25+213055.0  &  21 10 53.250  &  21 30 55.05  &  0.3657  &  LSlit   &  absorption lines only 									& 23.12\,$\pm$\,0.03	& 20.84\,$\pm$\,0.03\\
J211053.41+212931.5  &  21 10 53.414  &  21 29 31.45  &  0.3640  &  2108m6  &  [O\,{\sc ii}],H$\beta$,[O\,{\sc iii}] 						&			&\\
J211053.65+213049.8  &  21 10 53.650  &  21 30 49.80  &  0.3678  &  LSlit   &  [O\,{\sc ii}] 									& 22.14\,$\pm$\,0.03	& 19.95\,$\pm$\,0.03\\
J211053.71+213107.0  &  21 10 53.709  &  21 31 07.02  &  0.3088  &  LSlit   &  [O\,{\sc ii}],H$\beta$,[O\,{\sc iii}] 						& 23.68\,$\pm$\,0.04	& 22.38\,$\pm$\,0.03\\
J211053.76+213110.2  &  21 10 53.760  &  21 31 10.21  &  0.3603  &  LSlit   &  absorption lines only 									& 21.63\,$\pm$\,0.03	& 19.38\,$\pm$\,0.03\\
J211053.79+213029.5  &  21 10 53.790  &  21 30 29.52  &  0.3580  &  2108m5  &  absorption lines only 									& 22.52\,$\pm$\,0.03	& 20.23\,$\pm$\,0.03\\
J211053.80+213024.7  &  21 10 53.799  &  21 30 24.72  &  0.3647  &  2108m7  &  [O\,{\sc ii}] 									& 22.12\,$\pm$\,0.03	& 20.08\,$\pm$\,0.03\\
J211053.92+212907.1  &  21 10 53.917  &  21 29 07.06  &  0.3658  &  2108m6  &  [O\,{\sc ii}],H$\beta$,[O\,{\sc iii}] 						&			&\\
J211054.02+212847.6  &  21 10 54.025  &  21 28 47.61  &  0.2436  &  2108m2  &  [O\,{\sc ii}],H$\beta$,[O\,{\sc iii}] 						&			&\\
J211054.11+213057.7  &  21 10 54.111  &  21 30 57.73  &  0.3643  &  LSlit   &  absorption lines only 									&			&\\
J211054.16+212810.4  &  21 10 54.163  &  21 28 10.37  &  0.1837  &  2108m5  &  [O\,{\sc ii}],H$\beta$,H$\alpha$,[N\,{\sc ii}],[S\,{\sc ii}] 			&			&\\
J211054.36+213016.5  &  21 10 54.356  &  21 30 16.51  &  0.3649  &  2108m5  &  absorption lines only 									& 22.35\,$\pm$\,0.03	& 20.02\,$\pm$\,0.03\\
J211054.42+212936.1  &  21 10 54.418  &  21 29 36.13  &  0.3917  &  2108m5  &  [O\,{\sc ii}],H$\beta$,[O\,{\sc iii}] 						&			&\\
J211054.61+213228.9  &  21 10 54.610  &  21 32 28.86  &  0.7982  &  2108m7  &  [O\,{\sc ii}],H$\delta$,H$\gamma$,H$\beta$,[O\,{\sc iii}] 			& 23.83\,$\pm$\,0.03	& 22.23\,$\pm$\,0.03\\
J211054.81+212929.5  &  21 10 54.809  &  21 29 29.52  &  0.3670  &  2108m5  &  [O\,{\sc ii}],H$\beta$ 								&			&\\
J211054.97+212804.2  &  21 10 54.969  &  21 28 04.25  &  0.3907  &  2108m5  &  absorption lines only 									&			&\\
J211054.98+213032.5  &  21 10 54.980  &  21 30 32.55  &  0.3652  &  LSlit   &  absorption lines only 									& 22.85\,$\pm$\,0.04	& 20.75\,$\pm$\,0.03\\
J211054.98+212947.2  &  21 10 54.984  &  21 29 47.20  &  0.4261  &  2108m2  &  [O\,{\sc ii}]									&			&\\
J211055.03+213102.4  &  21 10 55.034  &  21 31 02.41  &  0.3581  &  LSlit   &  absorption lines only 									& 22.80\,$\pm$\,0.03	& 20.62\,$\pm$\,0.03\\
J211055.15+213115.1  &  21 10 55.152  &  21 31 15.12  &  0.3611  &  2108m2  &  [O\,{\sc ii}] 									& 23.52\,$\pm$\,0.03	& 21.85\,$\pm$\,0.03\\
J211055.34+212804.5  &  21 10 55.340  &  21 28 04.53  &  0.3910  &  2108m1  &  [O\,{\sc ii}],H$\beta$,[O\,{\sc iii}] 						&			&\\
J211055.71+213127.7  &  21 10 55.708  &  21 31 27.71  &  0.3618  &  2108m5  &  [O\,{\sc ii}],H$\beta$ 								& 22.52\,$\pm$\,0.03	& 20.35\,$\pm$\,0.03\\
J211055.82+213352.5  &  21 10 55.815  &  21 33 52.52  &  0.3600  &  2108m2  &  [O\,{\sc ii}] 									&			&\\
J211055.84+212926.3  &  21 10 55.843  &  21 29 26.32  &  0.6099  &  2108m4  &  absorption lines only 									&			&\\
J211055.84+213327.0  &  21 10 55.844  &  21 33 27.00  &  0.3612  &  2108m6  &  [O\,{\sc ii}],[O\,{\sc iii}] 							&			&\\
J211055.92+212921.7  &  21 10 55.922  &  21 29 21.70  &  0.8485  &  2108m5  &  [O\,{\sc ii}] 									&			&\\
J211056.00+213347.3  &  21 10 55.997  &  21 33 47.34  &  0.3609  &  2108m5  &  [O\,{\sc ii}] 									&			&\\
J211056.02+212803.1  &  21 10 56.019  &  21 28 03.11  &  0.3533  &  2108m7  &  [O\,{\sc ii}],H$\beta$,[O\,{\sc iii}] 						&			&\\
J211056.09+212816.4  &  21 10 56.092  &  21 28 16.42  &  0.3652  &  2108m7  &  [O\,{\sc ii}],H$\beta$,[O\,{\sc iii}] 						&			&\\
J211056.13+213432.8  &  21 10 56.133  &  21 34 32.77  &  0.2991  &  2108m5  &  absorption lines only 									&			&\\
J211056.35+213413.5  &  21 10 56.354  &  21 34 13.53  &  0.7962  &  2108m6  &  [O\,{\sc ii}] 									&			&\\
J211056.40+213218.6  &  21 10 56.400  &  21 32 18.65  &  0.3630  &  2108m6  &  [O\,{\sc ii}] 									& 20.74\,$\pm$\,0.03	& 18.73\,$\pm$\,0.03\\
J211056.62+213401.9  &  21 10 56.618  &  21 34 01.93  &  0.3625  &  2108m4  &  absorption lines only 									&			&\\
J211056.82+213158.9  &  21 10 56.824  &  21 31 58.92  &  0.3634  &  2108m5  &  absorption lines only 									& 22.88\,$\pm$\,0.03	& 20.60\,$\pm$\,0.03\\
J211056.86+213452.7  &  21 10 56.862  &  21 34 52.73  &  0.3639  &  2108m6  &  [O\,{\sc ii}] 									&			&\\
J211056.98+212918.5  &  21 10 56.978  &  21 29 18.54  &  0.3925  &  2108m7  &  [O\,{\sc ii}],H$\beta$,[O\,{\sc iii}] 						&			&\\
J211057.14+212749.4  &  21 10 57.137  &  21 27 49.41  &  0.6838  &  2108m6  &  [O\,{\sc ii}],H$\beta$ 								&			&\\
J211057.47+212759.8  &  21 10 57.473  &  21 27 59.80  &  0.3537  &  2108m1  &  [O\,{\sc ii}],[O\,{\sc iii}] 							&			&\\
J211057.50+213349.4  &  21 10 57.501  &  21 33 49.44  &  0.3628  &  2108m6  &  [O\,{\sc ii}] 									&			&\\
J211057.52+213335.0  &  21 10 57.521  &  21 33 35.04  &  0.3619  &  2108m5  &  [O\,{\sc ii}] 									&			&\\  \hline
\end{tabular}
\end{center}
\end{table*}

\begin{table*}
\begin{center}
\contcaption{}
\begin{tabular}{cccccccc} \hline
Object               & RA             & Dec           & Redshift & Mask     & Emission line(s) 									& F555W			& F814W \\ 
		        & (J2000)	  & (J2000)     &                 &               &                                                                                                                &                          &               \\  \hline
J211057.55+213310.5  &  21 10 57.547  &  21 33 10.54  &  0.3651  &  2108m6  &  [O\,{\sc ii}],H$\beta$,H$\alpha$ 						&			&\\
J211057.58+213402.7  &  21 10 57.575  &  21 34 02.71  &  0.3657  &  2108m6  &  [O\,{\sc ii}] 									&			&\\
J211057.61+213156.0  &  21 10 57.611  &  21 31 56.04  &  0.3680  &  2108m6  &  absorption lines only 									& 23.31\,$\pm$\,0.03	& 21.10\,$\pm$\,0.03\\
J211057.64+213255.8  &  21 10 57.635  &  21 32 55.76  &  0.3566  &  2108m5  &  absorption lines only 									&			&\\
J211057.80+213134.6  &  21 10 57.803  &  21 31 34.61  &  0.5125  &  2108m6  &  [O\,{\sc ii}],H$\beta$,[O\,{\sc iii}] 						& 23.98\,$\pm$\,0.04	& 22.03\,$\pm$\,0.03\\
J211057.82+213049.2  &  21 10 57.823  &  21 30 49.22  &  0.3629  &  2108m7  &  [O\,{\sc ii}],H$\beta$,[O\,{\sc iii}],H$\alpha$,[N\,{\sc ii}],[S\,{\sc ii}] 	& 23.02\,$\pm$\,0.03	& 21.46\,$\pm$\,0.03\\
J211057.83+213043.7  &  21 10 57.827  &  21 30 43.69  &  0.8252  &  2108m6  &  absorption lines only 									& 24.68\,$\pm$\,0.06	& 21.50\,$\pm$\,0.03\\
J211058.00+213144.9  &  21 10 58.003  &  21 31 44.89  &  0.3588  &  2108m5  &  absorption lines only 									& 22.91\,$\pm$\,0.03	& 20.66\,$\pm$\,0.03\\
J211058.08+213305.9  &  21 10 58.081  &  21 33 05.92  &  0.3585  &  2108m5  &  absorption lines only 									&			&\\
J211058.56+212902.8  &  21 10 58.558  &  21 29 02.76  &  0.3643  &  2108m6  &  absorption lines only 									&			&\\
J211058.56+213035.0  &  21 10 58.560  &  21 30 35.01  &  0.3662  &  2108m6  &  absorption lines only 									&			&\\
J211058.71+213326.9  &  21 10 58.715  &  21 33 26.94  &  0.1682  &  2108m4  &  [O\,{\sc ii}],H$\alpha$,[N\,{\sc ii}],[S\,{\sc ii}] 				&			&\\
J211058.92+213241.0  &  21 10 58.918  &  21 32 40.97  &  0.3607  &  2108m5  &  absorption lines only 									&			&\\
J211059.10+213251.1  &  21 10 59.096  &  21 32 51.15  &  0.3591  &  2108m5  &  absorption lines only 									&			&\\
J211059.60+213019.3  &  21 10 59.601  &  21 30 19.29  &  1.4815  &  2108m6  &  absorption lines only 									&			&\\
J211059.82+213104.7  &  21 10 59.816  &  21 31 04.74  &  0.3649  &  2108m6  &  absorption lines only 									& 22.84\,$\pm$\,0.03	& 20.40\,$\pm$\,0.03\\
J211059.86+213236.1  &  21 10 59.865  &  21 32 36.08  &  0.3599  &  2108m5  &  absorption lines only 									&			&\\
J211100.08+213254.1  &  21 11 00.085  &  21 32 54.13  &  0.3168  &  2108m6  &  [O\,{\sc ii}] 									&			&\\
J211100.67+213157.9  &  21 11 00.674  &  21 31 57.93  &  0.3639  &  2108m6  &  [O\,{\sc ii}] 									&			&\\
J211100.73+212919.8  &  21 11 00.732  &  21 29 19.76  &  0.5123  &  2108m6  &  [O\,{\sc ii}] 									&			&\\
J211100.86+212838.0  &  21 11 00.859  &  21 28 38.04  &  0.6456  &  2108m6  &  [O\,{\sc ii}],H$\beta$,[O\,{\sc iii}] 						&			&\\
J211100.98+212825.6  &  21 11 00.975  &  21 28 25.64  &  0.8275  &  2108m6  &  [O\,{\sc ii}] 									&			&\\
J211100.98+213249.7  &  21 11 00.984  &  21 32 49.68  &  0.0837  &  2108m4  &  H$\alpha$,[S\,{\sc ii}] 								&			&\\
J211103.21+213020.7  &  21 11 03.213  &  21 30 20.68  &  0.5880  &  2108m1  &  [O\,{\sc ii}] 									&			&\\
J211103.33+212938.3  &  21 11 03.331  &  21 29 38.34  &  0.3665  &  2108m4  &  [O\,{\sc ii}],H$\beta$,[O\,{\sc iii}] 						&			&\\
J211104.15+212856.7  &  21 11 04.150  &  21 28 56.72  &  0.2314  &  2108m2  &  [O\,{\sc ii}],H$\gamma$,[O\,{\sc iii}],H$\alpha$,[N\,{\sc ii}],[S\,{\sc ii}] 	&			&\\
J211104.21+212912.4  &  21 11 04.206  &  21 29 12.42  &  0.2325  &  2108m6  &  [O\,{\sc ii}],H$\beta$,[O\,{\sc iii}],H$\alpha$,[N\,{\sc ii}],[S\,{\sc ii}]	&			&\\
\hline
\end{tabular}
\end{center}
\end{table*}

\subsection{The lensing galaxies}

\begin{figure*}
\begin{center}
\setlength{\unitlength}{1cm}
\begin{picture}(12,6.5)
\put(-2.9,0.0){\includegraphics{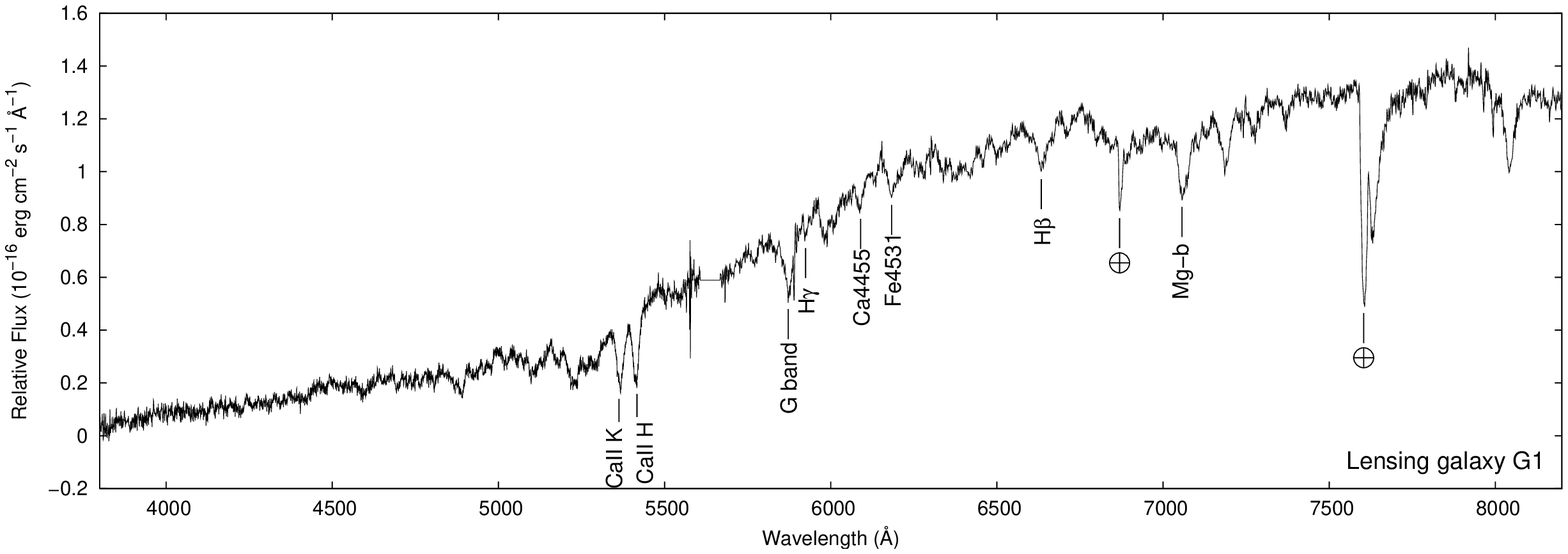}}
\end{picture}
\caption{The optical spectrum of the main lensing galaxy (G1) of B2108+213 taken with LRIS on the W. M. Keck Telescope on 2003 October 24. G1 is an early-type galaxy at a redshift of 0.3648\,$\pm$\,0.0002. The main absorption features used to determine the redshift of G1 have been identified and the skylines have been labeled with the crossed circles.}
\label{G1spec}
\end{center}
\end{figure*}

Our optical spectrum of the main lensing galaxy G1 is shown in Fig.~\ref{G1spec}. The spectrum has a spectral shape and several absorption features which are consistent with an early-type galaxy with an old (red) stellar population. The redshift of G1, established from eight absorption features (identified in Fig.~\ref{G1spec}), is 0.3648\,$\pm$\,0.0002. There is known to be a powerful radio-loud AGN embedded within G1 with radio-lobes which extend out to $\sim$~30~kpc from the core at 1.4~GHz \citep{more07}. Interestingly, there is no evidence of AGN activity in the form of emission lines, or a strong blue continuum in the spectrum of G1. Therefore, we assume that the AGN is either obscured at optical wavelengths or is currently going through a period where it is optically passive.

\begin{figure}
\begin{center}
\setlength{\unitlength}{1cm}
\begin{picture}(6,5.5)
\put(-1.3,0){\includegraphics{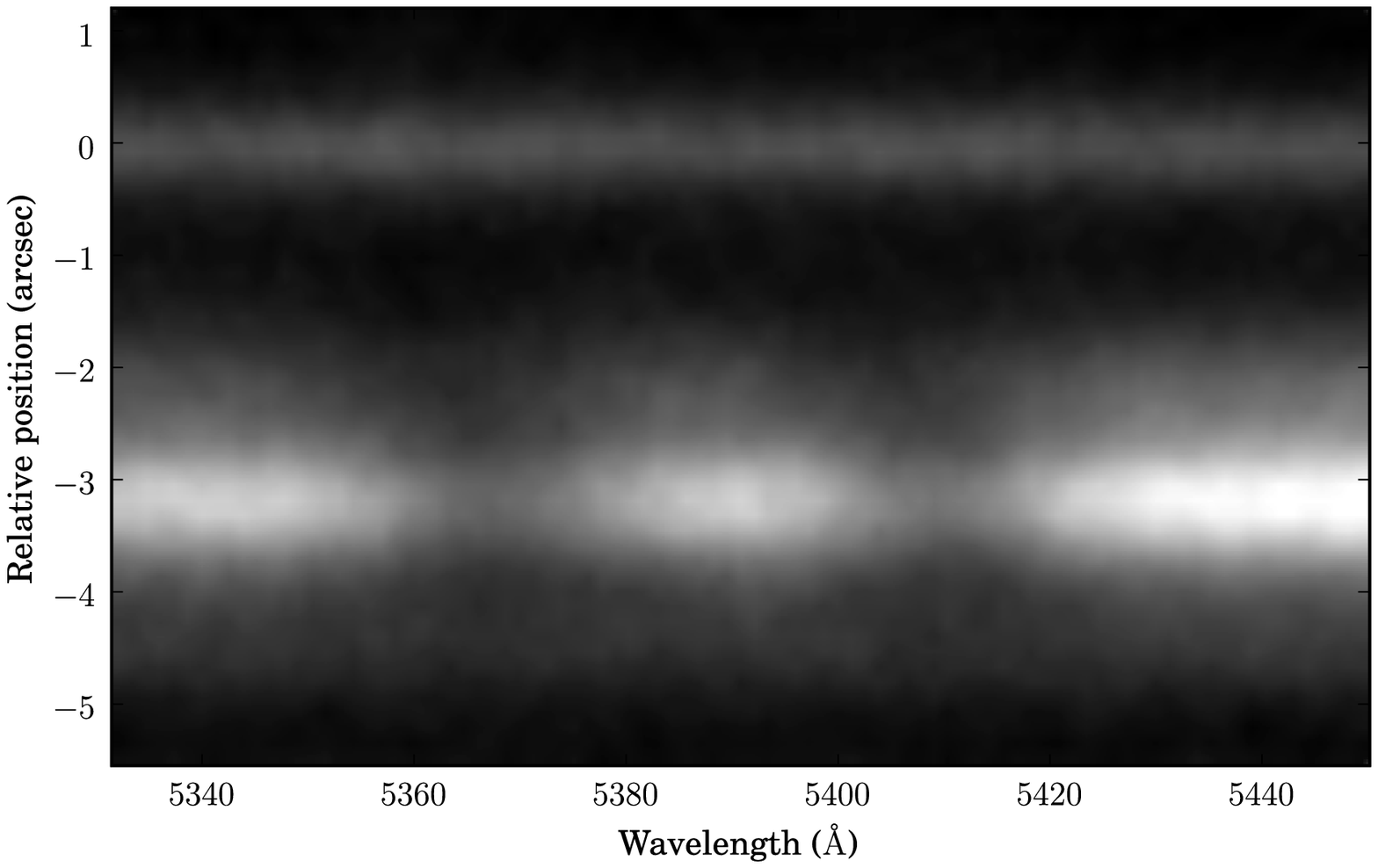}}
\end{picture}
\caption{The two dimensional spectrum of the B2108+213 lens system between 5330--5450~{\AA}. The position of lensed image A is marked as the origin of the spatial axis. The spectrum shows the clear Ca\,{\sc ii} H and K absorption lines in lensing galaxy G1 ($-$3.2~arcsec from lensed image A). The spectrum at the position of the second lensing galaxy ($-$2.2~arcsec from image A) shows a hint of the same Ca\,{\sc ii} H and K absorption lines. However, the signal-to-noise ratio and spatial resolution of the spectrum is not good enough to confirm that G2 is at the same redshift as G1.}
\label{G2plot}
\end{center}
\end{figure}

In Fig.~\ref{G2plot} we present part of the two dimensional spectrum from the slit placed over the second lensing galaxy, G2, during the multiple-slit spectroscopic observations on 2005 August 1 (4.5~h total integration time). It is clear that the 2-d spectrum is dominated by the much more luminous and extended galaxy G1. There is no evidence of any absorption lines over the full wavelength range of the spectrum that are at a different redshift from G1, but we do see absorption lines at a similar redshift to G1 in the G2 spectrum (see Fig.~\ref{G2plot} for an example of Ca {\sc ii} H and K absorption in both spectra). This suggests that G2 could be at the same redshift as G1. However, we cannot rule out that there is contamination of the weaker G2 spectrum by the spectrum of G1. Further observations, which spatially resolve G1 and G2, will need to be carried out to confirm that G2 is a companion to G1.

\subsection{The lensed source}

\begin{figure*}
\begin{center}
\setlength{\unitlength}{1cm}
\begin{picture}(12,6.5)
\put(-2.9,0.0){\includegraphics{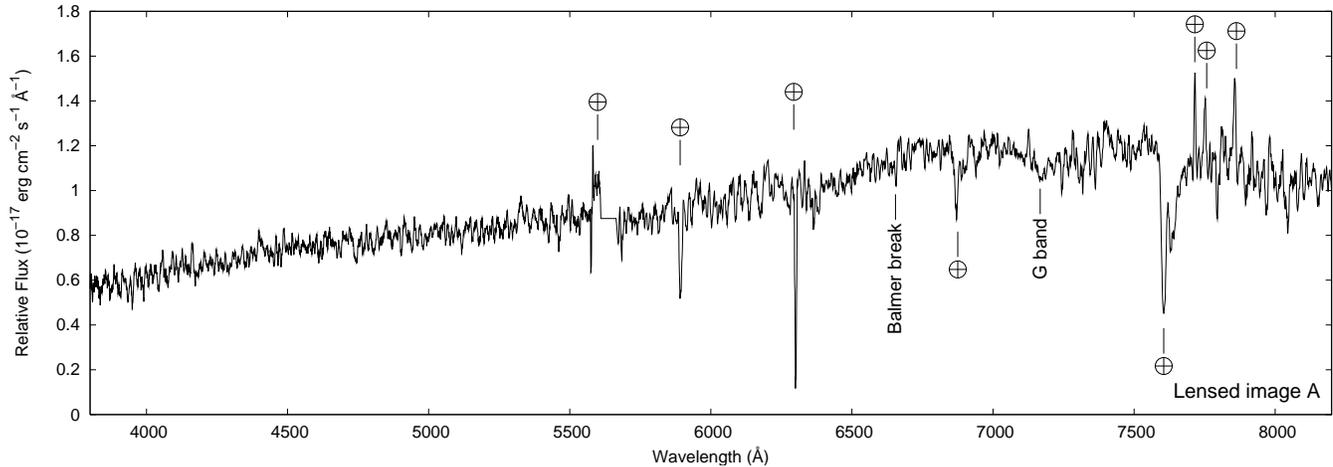}}
\end{picture}
\caption{The optical spectrum of the lensed quasar (image A) of B2108+213 taken with LRIS on the W. M. Keck Telescope on 2003 October 24. The flat optical spectrum of image A is consistent with a BL Lac type radio source. The red end of the spectrum shows some emission from the quasar host galaxy and contamination from the lens galaxy, G1. Sky emission and absorption lines have been labeled with the crossed circles. The spectrum has been smoothed using a boxcar of 7~{\AA}.}
\label{Aspec}
\end{center}
\end{figure*}

\begin{figure}
\begin{center}
\setlength{\unitlength}{1cm}
\begin{picture}(6,6.12)
\put(-1.3,0){\includegraphics{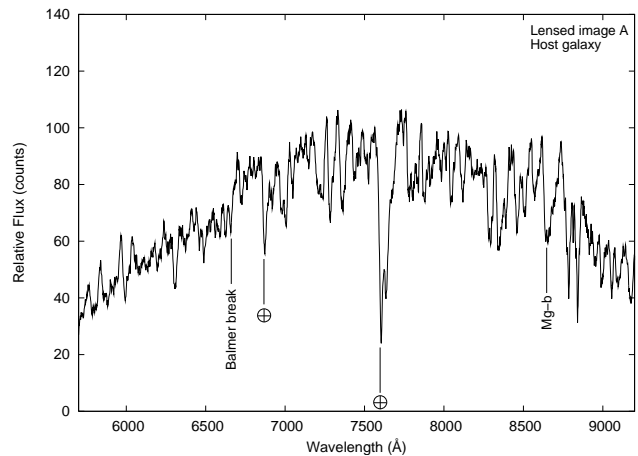}}
\end{picture}
\caption{The optical spectrum (5700--9200~{\AA}) of the gravitational arc from the lensed image A. The spectrum shows a possible spectral break and Mg-b absorption, giving a likely source redshift of 0.674\,$\pm$\,0.005. The spectrum has been smoothed with a box-car of width 20~{\AA} and has not been flux-calibrated or corrected for the spectral response of the instrument.}
\label{Aspec2}
\end{center}
\end{figure}

In Fig.~\ref{Aspec} we present the quasar spectrum of lensed image A. The blue-side of the spectrum is featureless and is slightly rising from $\sim$3800 to 5800~{\AA}. This spectral shape is consistent with a BL Lac type radio source (e.g., see \citealt{marcha96}), whose optical output is dominated by the non-thermal emission from the beamed quasar. The flat radio spectra of the lensed images of B2108+213, and the lack of extended structure observed in the radio maps \citep{mckean05,more07} are both in agreement with the classification as a beamed radio source viewed down the axis of the radio-jet. The red side of the spectrum shows a hint of emission from the quasar host galaxy, but also has contamination from the wings of lensing galaxy G1 (see Fig.~\ref{G1spec}). In Fig.~\ref{Aspec2} we show the spectrum of the extended arc of the host galaxy of the lensed quasar -- note that this spectrum has not been flux-calibrated, so the data are shown in relative counts without the shape of the bandpass being corrected for. There is evidence of a break in the spectrum at $\sim$6700~{\AA}, which we identify as the Balmer break. Note that the break in the spectrum cannot be the Lyman break since there is no evidence of absorption in the quasar spectrum below 6700~{\AA} by the Lyman forest.  We also see a step in the spectrum of the lensed quasar at 6660~{\AA} in the response corrected spectrum of image A shown in Fig.~\ref{Aspec}. Consistent with the Balmer break in the spectrum is the tentative detection of the Mg-b absorption line at 8661\,$\pm$\,27~{\AA} in Fig.~\ref{Aspec2}. This gives a likely redshift of $z_{s}=$~0.674\,$\pm$\,0.005 for the lensed source. Interestingly, there is an absorption line seen in the lensed quasar spectrum at 7172\,$\pm$\,17~{\AA} which could correspond to the G-band at redshift 0.666\,$\pm$\,0.004. However, this line is very close to a deep absorption line seen in the lens galaxy spectrum, which could be contaminating the absorption seen in the lensed quasar spectrum. There is no evidence of absorption line systems in any part of the spectrum of the lensed quasar, which would agree with the source redshift being relatively low.

\subsection{Large-scale structure velocity dispersion}

\begin{figure}
\begin{center}
\setlength{\unitlength}{1cm}
\begin{picture}(6,6.8)
\put(-1.2,-0){\includegraphics{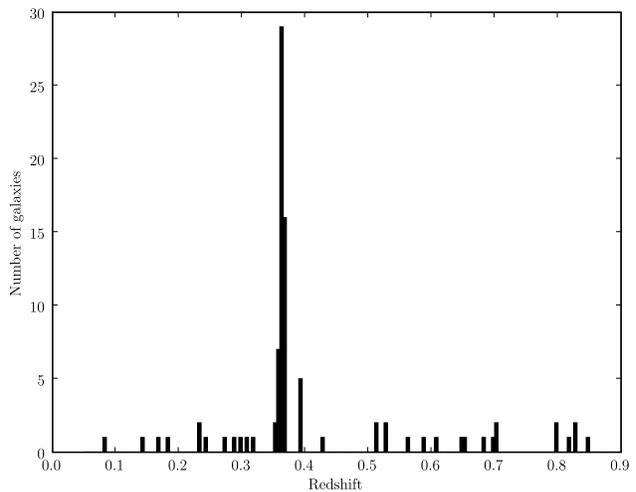}}
\end{picture}
\caption{The galaxy redshift distribution from the long and multi-slit datasets of the B2108+213 field. Redshifts were obtained for 90 extragalactic objects.}
\label{redhist}
\end{center}
\end{figure}

The redshift distribution of the 90 extragalactic objects found from the spectroscopic observations is shown in Fig.~\ref{redhist}. There is a clear spike at redshift 0.365, the redshift of the lensing galaxy G1. The galaxy membership of the large-scale structure was found by applying a linking velocity kernel. We define the linking velocity as an initial guess at the velocity dispersion of the structure. This linking velocity is also used to estimate the structure membership; the candidate galaxies are ordered by velocity and each member must be within the linking velocity of the next member. The group membership is refined using the iterative procedure outlined in \citet{wilman05} and \citet{auger07}. Due to the moderately large number of structure members, the bi-weight estimate of scale is used to determine the effective group velocity dispersion (see \citealt*{beers90}). Several linking velocities were used to estimate the group membership (450, 650 and 1100~km\,s$^{-1}$; see Table~\ref{group-vel}), which reaches a maximum of 52 members using a linking velocity of 1100 km\,s$^{-1}$. The three linking velocities returned group velocity dispersions of 451, 489 and 694~km\,s$^{-1}$, respectively, and in each case the velocity distribution is non-Gaussian. This suggests that the structure is not viralised and is still in the process of forming.

\begin{figure}
\begin{center}
\setlength{\unitlength}{1cm}
\begin{picture}(6,8.0)
\put(-1.2,-0){\includegraphics{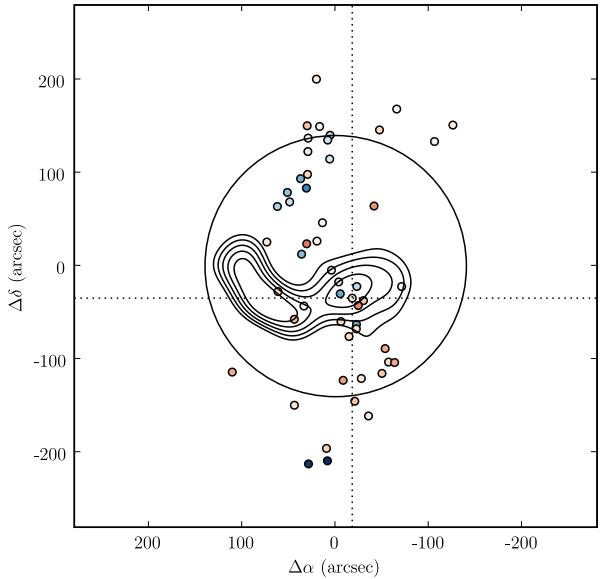}}
\end{picture}
\caption{The position of the group galaxies relative to the group centre for the 1100~km\,s$^{-1}$ linking kernel. The position of the lensing galaxy (G1) is shown by the cross. The intensity of the blue/red colour of each galaxy is dependent on the velocity relative to the systemic velocity of the group. The contours show the distribution of the X-ray emission around the lensing group [see \citet{fassnacht08} for details] and the circle marks a radius of 715 kpc.}
\label{positions}
\end{center}
\end{figure}

\begin{figure}
\begin{center}
\setlength{\unitlength}{1cm}
\begin{picture}(6,6.5)
\put(-1.6,-0.2){\includegraphics{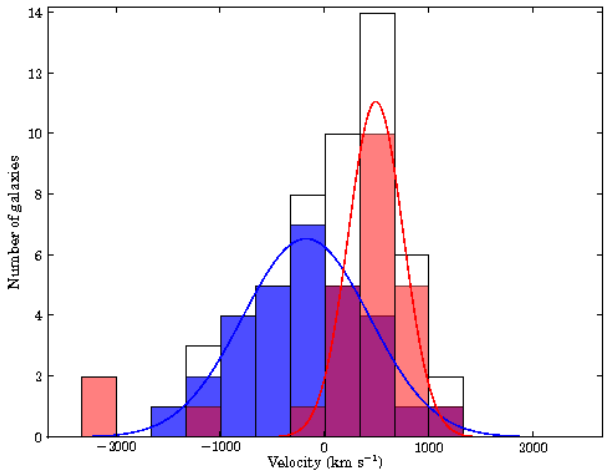}}
\end{picture}
\caption{The velocity distribution for the B2108+213 large-scale structure using a linking velocity kernel of 1100~km\,s$^{-1}$ (white). The velocity distributions of the blue and redshifted structures is also shown.}
\label{velocity}
\end{center}
\end{figure}

In Fig.~\ref{positions}, we show the positions of the 52 galaxies identified using the 1100~km\,s$^{-1}$ linking kernel. Also, the colour of each galaxy indicates how much the galaxy is blue or redshifted with respect to the systemic velocity of the structure. Interestingly, there appears to be an over-density of blueshifted galaxies to the northeast of the lens galaxy, and an over-density of redshifted galaxies to the southwest. Also, the divide between these blue and redshifted components is straddled by the X-ray emission detected with {\it Chandra} \citep{fassnacht08}. There is also no evidence of X-ray emission across the spatial distribution of the spectroscopically confirmed members of the lensing structure. It may be that we are observing the ongoing merger of two independent structures in this system. Based on this conclusion, we have identified two separate groups of galaxies; one which is north of the X-ray emission, and the other to the south. The velocity distributions of these two groups are shown in Fig.~\ref{velocity} as the blue and red distributions. Further observations will certainly have to be carried out to investigate the possibility that there are two merging groups in this system. For example, our redshift data at present are biased along an almost north-south orientation, with the lens system at the centre. This is because our imaging data with LRIS are orientated in this direction (see Table~\ref{log}).

\begin{table}
\begin{center}
\caption{The properties of the galaxy group associated with B2108+213 using a kernel of width 450, 650 and 1100 km\,s$^{-1}$. The right ascension and declination of the galaxy group are calculated from the mean position of the group galaxies.}\begin{tabular}{lcccc} \hline
Kernel & RA           & Dec         & Number of & $\sigma_v$ \\
       & (J2000)      & (J2000)     & galaxies  & (km\,s$^{-1}$) \\ \hline
450    & 21 10 55.147 & 21 31 36.92 & 41        & 451\,$\pm$\,45 \\
650    & 21 10 55.108 & 21 31 37.64 & 44        & 489\,$\pm$\,47 \\
1100   & 21 10 55.382 & 21 31 32.09 & 52        & 694\,$\pm$\,93 \\ \hline
\end{tabular}
\label{group-vel}
\end{center}
\end{table}

\section{Stellar kinematics and morphology of G1}
\label{kinematics}

The long-slit spectrum of G1 was taken with a 1~arcsec wide slit over an integration time of 1.5~h. This set-up produced a spectrum with a high enough spectral resolution ($\sim$174~km~s$^{-1}$ FWHM) and signal-to-noise ratio ($\sim$80) for the line-of-sight stellar velocity dispersion of G1 to be measured. The stellar velocity dispersion provides an additional constraint to the mass model and allows the average logarithmic matter density profile of the lensing galaxy to be measured when combined with the mass estimate from gravitational lensing (e.g. \citealt{treu04,sand04}; \citealt{koopmans06}).

The stellar velocity dispersion of G1 was measured using template spectra of eight G and K type stars \citep{sand04}. The templates were first smoothed to the resolution of the G1 spectrum, before the best-fitting velocity dispersion was determined from a linear combination of a polynomial continuum and the ensemble of templates. The measured stellar velocity dispersion within an aperture of 0.5~arcsec diameter is $\sigma_{\rm ap} =$~325\,$\pm$\,25~km\,s$^{-1}$. This is much larger than the characteristic velocity dispersion for the early-type lens galaxy population determined from gravitational lensing statistics ($\sigma^*=~$198$^{+53}_{-37}$~km~s$^{-1}$; \citealt{chae02}) and from the velocity dispersion function of galaxies \citep{sheth03}. However, the large value is typical of the most massive early-type galaxies that are usually found at the centres of groups and clusters (e.g. \citealt{bernardi07}).

The analysis of the high resolution optical and infrared imaging of B2108+213 with the {\it HST} found that the surface brightness distribution of G1 was well represented by a de Vaucouleurs profile \citep{mckean05}. However, fitting a single de Vaucouleurs profile to galaxy G1 revealed asymmetric residuals that suggested a more complex model for the light distribution was required. The data presented here have shown G1 to be part of a group or cluster, that it is the brightest group/cluster galaxy and has a large central stellar velocity dispersion. Also, {\it Chandra} imaging showed that G1 is close to the centre of the extended X-ray emission from the group/cluster \citep{fassnacht08}. Thus, it is likely that G1 is a giant elliptical cD galaxy (e.g. \citealt{matthews64}). Therefore, we have reanalysed the surface brightness profile of G1, using a new model that takes into account a large central bulge and an extended stellar envelope, typical of cD galaxies. This reanalysis is required because the effective radius of the bulge surface brightness profile is used in the dynamical models for the mass distribution in Section \ref{modelling}.

The surface brightness profile fitting of the lensing galaxy light from the {\sc acs} F814W and {\sc nicmos} F160W imaging was carried out using {\sc galfit} \citep{peng02}. The bulge of lensing galaxy G1, and all of the light from lensing galaxy G2 were fitted using S{\' e}rsic profiles, with the index set to $n =$~4 (de Vaucouleur models). The extended stellar envelope of G1 was fitted using a S{\' e}rsic profile, with the index left as a free parameter. The profiles were convolved with the {\it HST} point spread function prior to fitting \citep{krist95}. The extended arc emission from the lensed images was masked out.

Including a stellar envelope component results in a significantly better fit to the observed surface brightness distribution. For the {\sc acs} F814W dataset, the reduced $\chi^2$ of the model with a single de Vaucouleurs profile fitted to G1 is 1.8, whereas also including an envelope in the model lowers the $\chi^2$ to 1.3. Note that fitting a two component model to G1 also removed the asymmetric residuals that were seen around the galaxy in the previous analysis by \citet{mckean05}. The fitted S{\' e}rsic slope of the stellar envelope is 1.4 and the position is offset from the stellar bulge of G1 by $\sim$1.1~arcsec to the east. These results are both consistent with the morphology of envelopes in cD galaxies, which are often flattened and offset from the bulge. The parameters of the fitted light distributions to the F814W data are given in Table~\ref{hst-fit-tab}. Similar results are found for the {\sc nicmos} F160W dataset. However, the values of the slope (0.7) and effective radius (2.1 arcsec) for the stellar envelope were significantly different because the field of view of {\sc nicmos} is too small to properly fit the extended envelope component of G1. We find that by using a two component model for G1, the effective radius of the bulge is decreased by 16 per cent, when compared to what was found from the single component model \citep{mckean05}.

\begin{table}
\begin{center}
\caption{The surface brightness distributions of the two galaxies within the Einstein radius (G1 and G2). Only the parameters for the F814W imaging are presented because the field of view of the {\sc acs} is large enough to include all of the light from the extended stellar envelope of G1.}
\begin{tabular}{llllr}	\hline
Parameter			& \multicolumn{2}{c}{G1}		& G2		& Unit		\\
					& bulge		& envelope		&		&			\\ \hline
Total magnitude		& 17.74		& 17.95		& 20.42	& Vega-mag	\\
Effective radius			& 1.17		& 6.56		& 0.24	& arcsec		\\
Slope				& 4.00		& 1.44		& 4.00	&			\\
Axial ratio				& 0.88		& 0.81		& 0.77	&			\\
Position angle			& 61			& 161		& 128	& degrees		\\ \hline
\end{tabular}
\label{hst-fit-tab}
\end{center}
\end{table}

\section{Gravitational lens mass model}
\label{modelling}

In \citet{more07} it was shown that a simple mass model consisting of two singular isothermal spheres (SIS) plus a moderate external shear ($\gamma \sim$~0.04) could fit the observed VLBI constraints from the B2108+213 lensed images reasonably well, without the need for a large group or cluster contribution to the local shear and/or convergence. Despite this success, the flux-ratio of the two lensed images A and B could only be well-fit if the density profile of G1 (and G2 to a lesser extent) was steepened to a logarithmic density slope of $\gamma'=$~2.45$^{+0.19}_{-0.18}$ (68 per cent confidence level). This is relatively steep, but not inconsistent with some of the systems found in the SLACS survey \citep{koopmans06,koopmans09} or, for example, with PG~1115+080, another lens system closely associated with a compact group (e.g. \citealt{treu02,momcheva06}).

Here, we first attempt to include the contribution of the brightest nearby group members of B2108+213 into the gravitational lensing mass model in order to assess their effect, before incorporating the stellar dynamics of G1 and the extended arc structure. We also assess the effect of including a mass-sheet to account for the group. In addition, we attempt more general models with shear and a constant convergence gradient ($d\kappa/dr$) over the extent of the lensed images. The combination of these two can closely mimic the general effects of an external group or cluster for which a more precise mass model has yet to be established. To assess the relative effects of G1 and G2, and that of the group/cluster, we study two classes of mass models, namely, (i) isothermal ($\gamma' =$~2) and (ii) power-law density profiles; in both cases we investigate spherical and elliptical models.

\subsection{Models including the population of sub-haloes}
\label{model-vlbi}

We include the field galaxies as follows. For both the isothermal and non-isothermal cases, we normalize the masses inside the Einstein radius of each group member (including G2) to that of G1, using the proportionality relation $M\propto L^{1/2}$, where $M$ is the mass within a fixed aperture for an isothermal mass model. This case is consistent with the Faber--Jackson relation \citep{faber76}. All galaxies (besides G2) are assumed to be spherical to a first approximation. We also only include in our models those group members within 45~arcsec of the lens system, since out to this distance, we have good photometry and positions for the group galaxies from the {\it HST} {\sc acs} data. This adds eleven galaxies to the mass models. Even though there are galaxies found farther from the lens system, their contribution to the overall convergence is considered to be neglible; recall that the convergence $\kappa \sim b_E / r$ for an isothermal mass model, where $r$ is the distance from the galaxy and $b_E$ is the Einstein radius. We use exactly the same constraints on the models as in \citet{more07}. As a sanity check, we have tested our code against their results [for which the code of \citet{keeton01} was used] and confirm all their findings and $\chi^2$ values to within the expected numerical errors. 

Adding the field contribution to the models in the form of independent sub-haloes, besides an overall lowering of the lens-strength of both G1 and G2 (see Table \ref{model-data}) and an increase of the external shear, does not alter any of the model parameters outside the error range. The isothermal model with shear (2SIS+shear+field) has a reduced $\chi^2 =$~7.0, which is almost identical to the model of \citet{more07} where the field contribution was not included (reduced $\chi^2 =$~6.9). Like before, the model fits the positions of the lensed images A and B well, but fails to recover the observed flux-ratio. Therefore, solely adding the contribution of the field galaxies as sub-haloes fails to reproduce the observed properties of the lensed images. The additional convergence of the field results in a lowering of the Einstein radii of G1 and G2 by about 15 per cent. 

\begin{figure}
\begin{center}
\setlength{\unitlength}{1cm}
\begin{picture}(6,8.1)
\put(-1.6,-0.3){\includegraphics{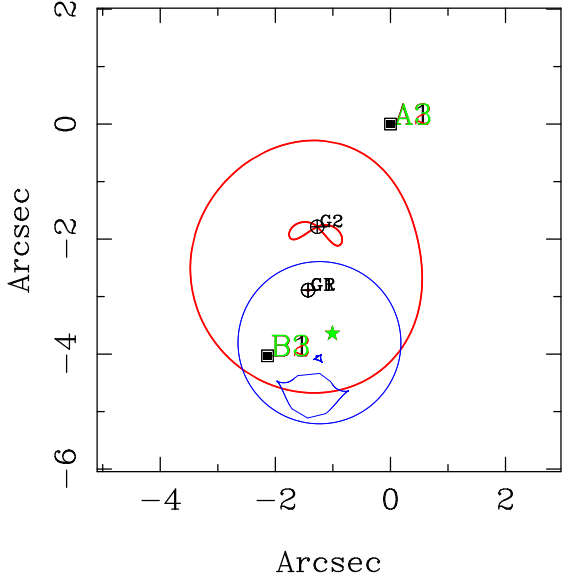}}
\end{picture}
\caption{The critical curves (red) and caustics (blue) of the mass model for B2108+213 that includes the field galaxies, an external shear and a convergence gradient.}
\label{model}
\end{center}
\end{figure}

We find that the positions and flux-ratios of the lensed images can both be reproduced if we invoke either a constant convergence gradient\footnote{The convergence gradient is defined as $d\kappa/dr$ and is added to the model at the position of the lensing galaxy, where its absolute value is 0. The strength of the gradient and its direction add two free parameters to the model.} of $d\kappa/dr =$~0.059\,$\pm$\,0.002~arcsec$^{-1}$ (2SIS+shear+field+gradient model), or a steeper density profile for the lensing galaxy G1 of $\gamma' =$~2.51\,$\pm$\,0.01 (2PLS+shear+field model) -- resulting in a reduced $\chi^2 \sim 0$ (see Table \ref{model-data}). However, it was not possible to differentiate between these two models based on the constraints provided by the VLBI observations of the lensed images. In Fig. \ref{model}, we show the critical curves and caustics for the model which includes an external shear, the field and a constant convergence gradient -- the critical curves for the other two models investigated here are similar. Note that the constant convergence gradient has a position angle of 15~degrees east of north (i.e. north-south), which is approximately increasing in the direction toward the centre of the X-ray emission found from this system \citep{fassnacht08}. 

\subsection{Models using lensing and stellar dynamics}
\label{model-velocity}

We now include in our analysis the stellar kinematic data to determine the average logarithmic density slope between the Einstein and effective radii of G1. The inputs to the kinematic model are the estimated stellar velocity dispersion (325\,$\pm$\,25 km\,s$^{-1}$) and the effective radius (1.17 arcsec) of G1. These models are constrained using the mass within the Einstein radius obtained from the gravitational lensing analysis. For the Einstein radius of G1, we have adopted a value of $b_E=$~1.51\,$\pm$\,0.26~arcsec, which was taken from the 2SIS+shear+field model (see Table \ref{model-data}). Note that we have increased the uncertainty on the Einstein radius of G1 to account for the possibility that G2 is more or less massive than has been assumed using the Faber-Jackson relation. The projected mass\footnote{The mass within the Einstein radius has been calculated assuming an isothermal mass distribution; $M_E =$~1.24\,$\times$\,10$^{11}$\,$b_E^2$\,($D_l D_s / D_{ls}$)~$M_\odot$, where $D_l$, $D_s$ and $D_{ls}$ are the angular diameter distances to the lens, source and between the lens and source, respectively.} within the Einstein radius of G1 is found to be 7.2\,$\pm$\,2.5\,$\times$\,10$^{11}$~$M_\odot$. Using this mass and its uncertainty, the spherical Jeans equations are solved following the method described in, for example, \citet{koopmans06}. We have also tested for the effect of anisotropy in the stellar orbits ($\beta = \pm$\,0.25) and of using either a Hernquist or Jaffe model for the stellar component of G1.

\begin{figure}
\begin{center}
\setlength{\unitlength}{1cm}
\begin{picture}(6,8.6)
\put(-1.5,-2.2){\includegraphics{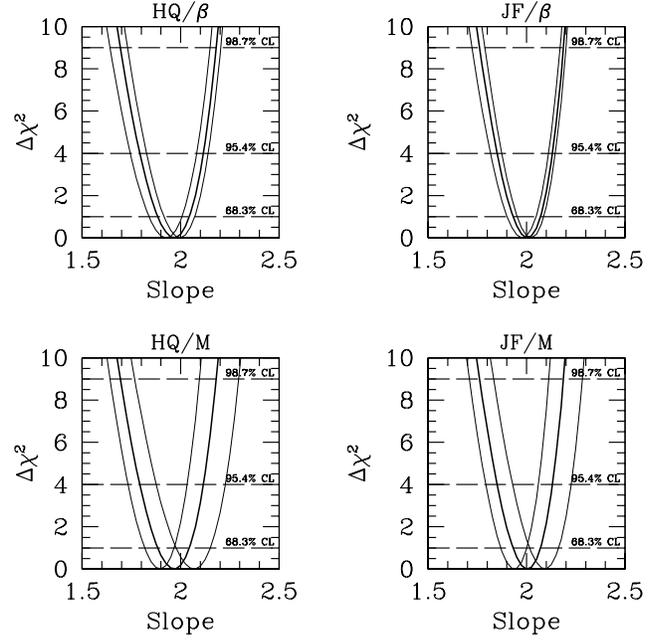}}
\end{picture}
\caption{The best fitting logarithmic density slope of the mass distribution for G1 from a joint lensing and dynamics analysis. ({\it Top}) The thick black lines are for an isotropic distribution of stellar orbits, whereas, the thin grey lines correspond to an anisotropy of $-$0.25 and +0.25. ({\it Bottom}) The thick black lines are for the mass of G1 assuming the Faber--Jackson relation holds for G2. The thin grey lines are for the cases were G2 is more or less massive than predicted from the Faber--Jackson relation. ({\it Left and right}) The analysis for Hernquist (HQ) and Jaffe (JF) models for the stellar mass distribution of G1.}
\label{dynamics}
\end{center}
\end{figure}

In Fig. \ref{dynamics}, we show the $\chi^2$--logarithmic density slope plots for lensing galaxy G1. For each of the combinations of mass, stellar model and anisotropy, the slope of the density profile is close to isothermal. We see that the largest scatter in the derived slope comes from the possible range of masses that G1 can have due to the uncertainty of including galaxy G2 in the model. The best fitting density slope is $\gamma' =$~1.98$^{+0.06}_{-0.10}$ for the Hernquist model and $\gamma' =$~2.00$^{+0.07}_{-0.07}$ for the Jaffe model (both at the 68 per cent confidence level). The range of slopes allowed for the possible masses that G1 can have is 1.82~$\leq \gamma' \leq$~2.16 (68 per cent confidence level). Therefore, it is clear from the lensing and dynamics analysis of the mass distribution presented here that the average logarithmic density slope of G1 is consistent with an isothermal mass model. We note that the combined luminous and dark matter density slope of $\gamma'=$~2.51 required to reproduce both the positions and flux-ratios of the lensed images by the lensing model alone would require the stellar velocity dispersion of G1 to be of order $\sim$~600~km\,s$^{-1}$.

\begin{table*}
\begin{center}
\caption{The parameters of the gravitational lens models for B2108+213. All positions are given relative to lensed image A1 and all position angles are measured east of north. The stated model parameters are the median values obtained by carrying out a MCMC simulation and the resulting uncertainties have been found from the 68 per cent confidence levels. Note that these are statistical uncertainties and do not reflect any uncertainties due to the choice of models used.}
\begin{tabular}{lllll}	\hline
Model							& 2SIS+shear+field				& 2SIS+shear+field+gradient				& 2PLS+shear+field			&		\\ \hline
G1 position ($\Delta\alpha$, $\Delta\delta$)		& 1431\,$\pm$\,1, $-$2888\,$\pm$\,1		& 1431\,$\pm$\,1, $-$2888\,$\pm$\,1		& 1431\,$\pm$\,1, $-$2888\,$\pm$\,1	& mas		\\
G1 Einstein radius					& 1511\,$\pm$\,6				& 1425\,$\pm$\,3				& 1893\,$\pm$\,11			& mas		\\
G1 slope						&						&						& 2.51\,$\pm$\,0.01			&		\\
G2 position ($\Delta\alpha$, $\Delta\delta$)		& 1273\,$\pm$\,5, $-$1786\,$\pm$\,5		& 1273\,$\pm$\,4, $-$1786\,$\pm$\,4		& 1273\,$\pm$\,5, $-$1786\,$\pm$\,5	& mas		\\
G2 Einstein Radius					& 437						& 415						& 549					& mas		\\
External shear						& 0.065\,$\pm$\,0.001				& 0.040\,$\pm$\,0.001				& 0.088\,$\pm$\,0.005			&		\\
External shear position angle				& 95\,$\pm$\,2					& 92\,$\pm$\,2					& 77\,$\pm$\,1				& degrees	\\
Gradient						& 						& 59\,$\pm$\,2					&					& mas$^{-1}$	\\
Gradient position angle					&						& 15\,$\pm$\,4					& 					& degrees	\\
Source 1 position ($\Delta\alpha$, $\Delta\delta$)	& 956, $-$3542					& 1010, $-$3646					& 741, $-$1957				& mas		\\
Source 1 flux-density					& 1.1						& 0.8						& 3.8					& mJy		\\
Source 2 position ($\Delta\alpha$, $\Delta\delta$)	& 955, $-$3542					& 1008, $-$3645					& 739, $-$1956				& mas		\\
Source 2 flux-density					& 0.4						& 0.3						& 1.3					& mJy		\\
Source 3 position ($\Delta\alpha$, $\Delta\delta$)	& 953, $-$3543					& 1007, $-$3645					& 736, $-$1958				& mas		\\
Source 3 flux-density					& 0.2						& 0.1						& 0.6					& mJy		\\
Flux-ratio (B1/A1)					& 0.63						& 0.45						& 0.46					&		\\
Reduced $\chi^2$ 					& 6.96						& 0.01						& 0.07					&		\\ \hline
\end{tabular}
\label{model-data}
\end{center}
\end{table*}

\subsection{Models using the extended arc emission}
\label{model-arc}

We now test models using the extended arc emission from the background quasar host galaxy that is seen in both lensed images. In recent years, sophisticated grid-based modeling algorithms have been developed, allowing the properties of the lensing potential to be found and for the background object to be reconstructed from the extended source intensity distribution (e.g. \citealt{warren03,koopmans05,suyu09,vegetti09}). Here, we attempt to constrain the radial profile of the lensing mass distribution by investigating the radial structure of the extended arc emission. For example, for an isothermal density profile, the radial stretching should be identical for both lensed images (e.g. \citealt*{wucknitz04}). The modelling was carried out with the adaptive and fully Bayesian grid-based code of \citet{vegetti09}. We use for our analysis the {\sc nicmos} F160W data for B2108+213 that has the light from the lensing galaxies removed and any residuals masked out. We have also masked the bright quasar emission for two reasons. First, the quasar emission only supplies limited constraints to the mass model while introducing a strong curvature in the reconstructed source. Second, in the modelling, we have used a truncated version of the psf, which typically produces negligible effects below the noise for most pixels in the image, but would produce non-negligible artifacts for those pixels containing the strong point-source quasar emission. Ideally we would like to differentiate between the isothermal models predicted from the lensing and dynamics analysis and the models with a steep power-law slope of $\gamma'\sim$~2.5 found from the VLBI lensing data alone. Therefore, we have used the Bayesian evidence values to estimate the best fitting model. 

\begin{figure*}
\begin{center}
\setlength{\unitlength}{1cm}
\begin{picture}(6,8.5)
\put(-5.7,-0.0){\includegraphics{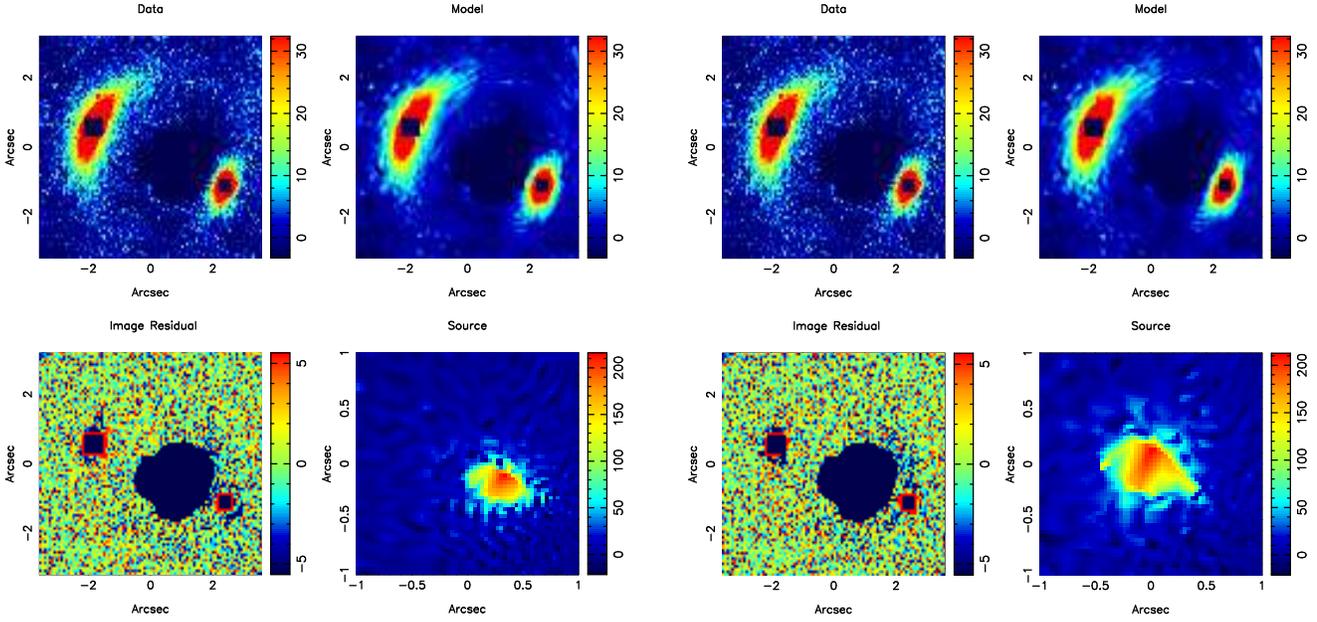}}
\end{picture}
\caption{The results of a grid-based modelling analysis of the extended arc emission from the B2108+213 host galaxy. ({\it Left}) The observed emission, modelled emission, image residuals and source reconstruction for an isothermal mass model. ({\it Right}) The same plots for a power-law model with a logarithmic density slope of $\gamma'=$~2.50. Note that the images are orientated as they were observed with the {\it HST}, that is, rotated by 96 degrees east of north.}
\label{grid.55}
\end{center}
\end{figure*}

In Fig.~\ref{grid.55}, we show the results of fitting an isothermal sphere mass model and a spherical power-law model with a density slope of $\gamma' =$~2.50. For the case with an isothermal mass model, the Einstein radii of lensing galaxies G1 and G2 are 1.87 and 0.39~arcsec, respectively. Note that for these models the mass-ratio of G1 and G2 has been left as a free parameter. The shear strength is found to be 0.03 at a position angle of 88~degr east of north. These model parameters are similar to, but not formally consistent with the results obtained using only the VLBI constraints from the lensed images \citep{more07}. The source has an elliptical shape with the peak surface brightness at the centre of the reconstructed host galaxy, which is plausible. For the spherical power-law model, the Einstein radii of G1 and G2 are lowered to 1.65 and 0.40~arcsec, respectively, and the shear is increased to 0.075 at a position angle of 82~degr east of north. Again, the source reconstruction looks plausible.

Both the isothermal and power-law models result in a good reproduction of the observed source intensity distribution (see Fig. \ref{grid.55}), although there are still some systematics left in the images. Formally, the power-law model is a better fit to the data, with a Bayesian evidence difference of $\Delta \log_{10} E =$~87 [see \citet{vegetti09} for a discussion on Bayesian evidence values and gravitational lens modelling]. Note that we did not include a convergence gradient for the isothermal case during this analysis because here we only wanted to differentiate between the two possible density slopes using the extended infrared emission from the lensed source.

\subsection{Models including a mass--sheet}

As a final test, we have included a flat mass-sheet into the lens model to account for any external convergence that is not taken care of through the inclusion of the satellite galaxies of the group. The effect of a mass-sheet would be to decrease the amount of mass attributed to G1 in the lens models, which would in turn result in the logarithmic density slope of G1 to increase for a fixed stellar velocity dispersion. This degeneracy between the slope of the density profile and the presence of an underlying invisible mass-sheet is often referred to as the mass-sheet degeneracy, and can have important implications for determining, for example, the Hubble constant from gravitational lenses (see e.g. \citealt{suyu10}). However, with the measurement of the stellar velocity dispersion of the galaxy, this degeneracy can be broken and the average slope of the inner density profile can in principle be determined (see Section \ref{model-velocity}). Alternatively, we can use the slope required from the gravitational lensing model and the measured stellar velocity dispersion to calculate the external convergence from the mass-sheet that is needed to reconcile the two. For this model, we use the density slope $\gamma'=$~2.45$^{+0.19}_{-0.18}$ determined from the VLBI constraints \citep{more07}, with a Gaussian prior on the slope. We also use the stellar velocity dispersion of 325\,$\pm$\,25~km\,s$^{-1}$ for G1, with a Hernquist profile for the galaxy light distribution. The break radius of the galaxy is set to 50~kpc, after which the logarithmic density slope becomes $\gamma'_{\rm outer}=$~3.

\begin{figure}
\begin{center}
\setlength{\unitlength}{1cm}
\begin{picture}(6,8.6)
\put(-1.8,-0.3){\includegraphics{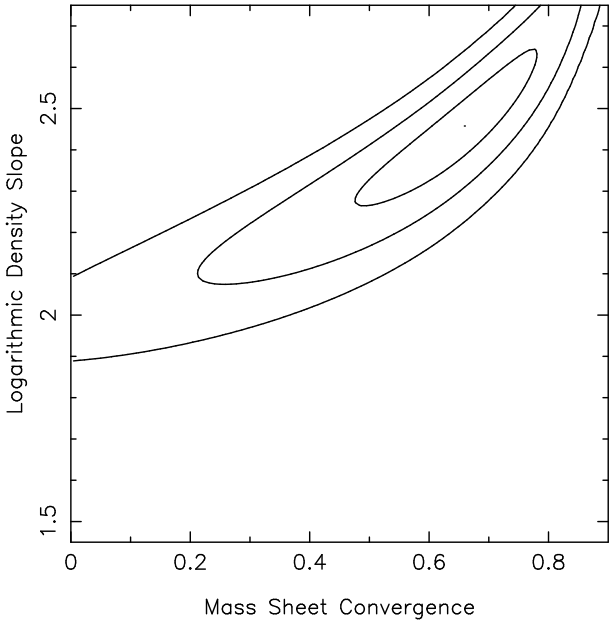}}
\end{picture}
\caption{The external convergence of the group halo as a function of the logarithmic density slope of G1. The contours shown are for 68, 95 and 99 per cent confidence limits. The most likely value of the external convergence for a density slope of $\gamma'=$~2.45$^{+0.19}_{-0.18}$ and a stellar velocity dispersion of 325\,$\pm$\,25~km\,s$^{-1}$ is $\kappa_{\rm ext} =$~0.65\,$\pm$\,0.15 (stat.)\,$\pm$\,0.15 (syst.).}
\label{external}
\end{center}
\end{figure}

The result from this model is presented in Fig.~\ref{external} where the inner logarithmic density slope of G1 is shown as a function of the external convergence. We see that increasing the external convergence results in a steeper inner density profile for G1, as expected. It is clear that we need a large mass sheet to explain the differences in the slope between using only the VLBI constraints and using the joint lensing and dynamics analysis. The most likely value for the external convergence from the mass-sheet is $\kappa_{\rm ext} =$~0.65\,$\pm$\,0.15 (stat.)\,$\pm$\,0.15 (syst.) to explain this difference, in units of critical density. This $\kappa_{\rm ext}$ for the mass-sheet is much higher than the value obtained from just including the group galaxies into the model as sub-haloes ($\kappa_{\rm ext} =$~0.12 at the position of image A\footnote{For the 2SIS+shear+field model. A similar value is obtained at the position of image B.}), which implies there would have to be a significant underlying parent halo to represent the group dark matter distribution. Such a large external convergence would be expected to produce further multiple imaging in the field of B2108+213. However, none is seen in the {\it HST} imaging.

\subsection{Possible systematics that could effect the modelling}

The lensing mass models investigated here have been constrained by the flux-ratios of the lensed images, leading to the deviation from isothermal profiles for G1 and G2. It is worth noting that the lensed source is a compact flat-spectrum radio source, which in the most luminous cases are typically variable across the whole spectrum. However, the flux-ratios of the two lensed images over optical and radio wavelengths, even when observed at different epochs, are almost identical \citep{mckean05,more07}. This suggests that source variability coupled with a lensing time-delay is not causing the flux-ratio between images A and B to differ from what is expected from an isothermal model. Furthermore, the lens models that use only the extended arc emission from the quasar host galaxy, whose emission is not expected to be variable, favour density profiles that are steeper than isothermal. The radio spectra of the two lensed images between 1.4 and 8.46 GHz are also similar, ruling out free-free absorption as a possible cause for the flux ratio not being as expected for a simple isothermal mass distribution (e.g. \citealt{mittal07}; \citealt*{winn03}).

Another, more likely mechanism that could be possibly responsible for changing the flux-ratio of the lensed images is mass substructure (e.g. \citealt{dalal02}). An inspection of the high resolution {\it HST} imaging of B2108+213 does not show any evidence for small-scale structure in the form of dwarf galaxies near to the lensed images. Such dwarf galaxies have recently been found to cause anomalous flux-ratios in some gravitational lens systems (e.g. \citealt{mckean07,more09,macleod09}). The radio emission from the lensed images shows some extended emission on VLBI scales, but it is not sufficient to investigate whether there is mass substructure near to the lensed images. However, the modelling of the lensed emission using the extended gravitational arc found that the arc structure could be well reproduced by a globally smooth mass model with no significant residuals (e.g. \citealt{koopmans05,vegetti09}).

The lensing and dynamics models rely on determining the mass enclosed within the Einstein radius, which is dependent on the redshifts of the lens and source being known. Although there is no doubt over the measured redshift of the lens, the source redshift is less secure since it relies on the detection of a break in the spectrum, as opposed to several definite detections of emission and absorption lines. It is worth noting that if the source redshift is higher than 0.67 then there will be little change in the enclosed mass due to the curvature in the angular diameter distances as a function of redshift. If the source redshift is lower, then the mass of the lens would need to be higher to explain the wide image separation, which would force the lensing and dynamics models to produce a flatter logarithmic density slope. Therefore, further observations to confirm the source redshift should be carried out to remove this possible error in the lensing and dynamics models.

\section{Discussion}
\label{disc}

B2108+213 is an interesting system because i) the wide separation between the two lensed images of 4.56~arcsec suggests a group-scale lens and ii) previous attempts to model the system have required that the main lensing galaxy has a steeper than isothermal density profile. We now discuss to what extent our results have explained these unusual features of this lens system and what it tells us about the mass distribution of a group-sized halo.

\subsection{Merging groups or a cluster?}

We first discuss whether the lens system lies in a group or a cluster. Our spectroscopic survey of the B2108+213 field determined redshifts for 90 galaxies. Of these galaxies, up to 52 (including the main lensing galaxy G1) were found to be at redshift $\sim$0.365. Thus, we have confirmed that there is a large-scale structure associated with this gravitational lens system. The total size of the structure is between 41 to 52 galaxies, depending on the linking kernel used to find the group membership. From Figs. \ref{cmVR} and \ref{cmRI} it is clear that there is a well defined red sequence in the large-scale structure. This is not too surprising considering that we preferentially selected red galaxies by their colour. Since we did not search for the less massive late-type galaxies during our survey it is likely that the galaxy membership of the large-scale structure could be much larger. The large number of members already confirmed would suggest that this structure has an Abell richness of 0 \citep{abell58}. The velocity dispersion of the galaxies is between 451 to 694~km\,s$^{-1}$ depending on the linking kernel used. This is potentially much larger than the typical velocity dispersions of low redshift groups, which tend to be around 200--450~km\,s$^{-1}$ \citep{zabludoff98}, and would point toward this structure being at most a poor cluster. However, we also found that the velocity distribution of the galaxies has a spatial structure, suggesting that there may be two galaxy groups in the process of merging to form a cluster. Further observations are required to confirm this.

The X-ray data for this system give the strongest indication that this structure is a group of galaxies. The X-ray luminosity of the diffuse intra-group gas is $L_{X,bol,500} \sim$ 1.2~$\times$~10$^{43}$~ergs~s$^{-1}$, which is toward the low end of the luminosity distribution for clusters at a similar redshift \citep{ebeling00,kocevski07}. Furthermore, the X-ray luminosity is over an order of magnitude lower compared to what is expected for intermediate redshift groups and clusters with velocity dispersions as large 694~km\,s$^{-1}$, the velocity disperion found for the largest linking kernel \citep{willis05,jeltema06,fassnacht08}. Note that for the 450 and 650 km\,s$^{-1}$ linking kernels, the group velocity dispersion is $\sim$\,470~km\,s$^{-1}$, placing the B2108+213 group on the $L_X$--$\sigma_v$ relation for intermediate redshift groups (see \citealt{fassnacht08}). The X-ray emission also does not extend over the spatial distribution of the galaxies in the structure, but it is almost centred on the position of the brightest group galaxy, lensing galaxy G1 -- the projected distance between the X-ray peak and lensing galaxy G1 is $\sim$36~kpc (see Fig. \ref{positions}; \citealt{fassnacht08}). Finally, we find that the main lensing galaxy is a cD galaxy, which must have formed through the merging of smaller mass structures. Therefore, we believe there is strong evidence that this is a dynamic multiple-group system in the process of merging.

There are a number of avenues which can be taken to explore the full extent of the B2108+213 large-scale structure in the future. The most useful would be to carry out a weak lensing analysis from wide field imaging of the system. This would show whether the structure has an extended dark matter halo, or possibly shows evidence for two or more merging haloes. It would also be interesting to see whether the dark matter halo follows the X-ray emission, or is closer to the emission from the galaxies -- as has been recently observed in merging clusters (e.g. \citealt{clowe06,bradac06,bradac08}). Attempts to measure a weak lensing signal from {\it HST} and ground-based data are currently underway.

\subsection{Mass distribution of a group-scale halo}

Having established that the image-splitting of B2108+213 is probably caused by a group-scale halo, we now discuss what the gravitational lensing and stellar dynamics data can tell us about the distribution of mass within the structure.

Including the contribution of the group, either as sub-haloes and/or as a mass gradient (or constant mass-sheet) to represent an underlying parent group halo, results in lowering the Einstein radii of the two lens galaxies between the lensed images, and hence their projected mass determined from gravitational lensing. This highlights the importance of carrying out large spectroscopic surveys of the environments of lens systems for calculating the projected mass of lensing galaxies. The total convergence due to the sub-halo population is $\kappa\sim$~0.12, which is higher than that found for other gravitational lens systems associated with groups and/or having groups found along the line-of-sight (e.g. \citealt{fassnacht06,momcheva06,auger07,treu09}). This is due to the other gravitational lens systems being part of loose groups of only a few members. For B2108+213, there are up to 52 confirmed group members. Also, from Fig. \ref{slit-pos} it is clear that there is a compact structure of 5 galaxies that are only 15~arcsec from the lens system, or at a projected distance of $\sim$75~kpc. Therefore, B2108+213 probes the mass distribution at the centre of a compact structure.

Our models using only the lensing data obtained from the high resolution imaging with VLBI or the {\it HST} strongly favour steeper than isothermal density profiles {\it at} the position of the Einstein radius, in the absence of some external convergence field. Such a steep density profile for the main lensing galaxy is certainly not unique (e.g. \citealt{koopmans06}) and has been predicted by simulations for haloes that have recently undergone some kind of interaction \citep*{dobke07}. However,  the steep density profile appears to be inconsistent with what was found when the stellar kinematics data are incorporated into the model, which requires the average logarithmic density slope between the Einstein and effective radius of G1 to be close to isothermal. A possible explanation for this contradiction is that the mass distribution of the main lensing galaxy can not be simply approximated by a single power-law, but by perhaps, a broken power-law with a knee somewhere between the effective radius and the Einstein radius of the system. There are two physical mechanisms that could produce such a broken power-law, which we now discuss.

First, the total mass density profile predicted from simulations is known not to be a single power-law at cluster scales (e.g. \citealt{sand04,sand08}); the innermost region is dominated by the brightest cluster galaxy stars, leading to a total mass density with a slope approximately isothermal or steeper. At larger radii the inner part of the dark matter halo takes over with a typical slope between 0 and 1. At even larger radii the profile becomes steeper -- in accordance with the predictions of numerical simulations, reaching again isothermality and then an even steeper slope of $\sim$~3 towards the virial radius (e.g. \citealt{kneib03}).  The so-called break radius, where the dark matter profile is approximately isothermal, is of order 100~kpc for clusters. In contrast, at the scale of massive galaxies, the smaller break radius and the high concentration of stars in the centre, conspire to create a total mass density profile that is consistent with a single power law over more than two decades in radius \citep{gavazzi07}. B2108+213 explores the intermediate regime of a group size halo, for which we expect the break radius to be of order of tens of kpc\footnote{For our models we assumed a break radius of 50 kpc.} \citep{maccio07}. Thus, we could interpret the change in the average slope between the Einstein and effective radius as a signature of a turnover in the mass density profile, qualitatively in agreement with simulations. However, to explain the flux-ratio of B2108+213 would require the turnover to occur close to the Einstein radius, that is, at around 8~kpc, which seems unlikely for a group-scale halo. More radially extended data would be needed to pinpoint the location of a possible break, the virial mass of the structure, and the actual slope in the outer parts. In a future paper, we plan to add weak lensing to our constraints to investigate the full extent of the radial profile.

A second mechanism that could account for the changing slope at the inner part of the halo is galaxy truncation. Simulations have shown that as galaxies fall into a massive potential, like in the cases of groups or clusters, the dark matter is removed in the outer parts of the galaxy's halo by tidal stripping (e.g. \citealt{bullock01,gao04,limousin09b}) leading to a truncated halo. Weak lensing measurements around galaxies in clusters have found evidence for halo truncation, with those galaxies closest to the centre of the cluster showing the largest levels of truncation \citep{gavazzi04,natarajan09}. The truncation radius of these galaxies is of order $\sim$~50--60~kpc (\citealt*{halkola07}; \citealt{natarajan09}). It would be reasonable to expect a high level of truncation for the lensing galaxy G1 -- the galaxy sits close to the centre of the group and is likely to have undergone many mergers and interactions over the course of time that the group has taken to form.

A problem with the broken power-law model is that it requires the inner part of the halo to be quite flat for the {\it average} logarithmic density profile to be close to isothermal. This would essentially need G1 to have a more centrally concentrated mass distribution. To test this, we considered a model with a constant mass-to-light ratio (as opposed to $M \propto L^{1/2}$ from the FJ-relation) for G1. We find that although the lensing constraints can still be fitted with this model, the velocity dispersion of G1 is pushed higher, to around 450 km\,s$^{-1}$ and above, since we must now have more mass within the effective radius of G1. This is clearly not consistent with the observed stellar velocity dispersion of G1. A possible solution is that there is some external convergence from the group that would result in G1 being less massive, but this again would need to be significant ($\kappa_{\rm ext} \sim$~0.6) and is not plausible. For this reason, it seems that some sort of broken power-law model is not likely.

Our simplest model for the mass distribution of the lens system that satisfies both the lensing and dynamics constraints, has an underlying external convergence gradient. Applying a convergence gradient is successful because it changes the expected flux-ratios of the lensed images, which from the data currently in hand appear not to have been strongly affected by mass substructure or by variability. This gives a lens model that is consistent with an isothermal mass distribution, and hence with the stellar velocity dispersion of G1. The moderate convergence gradient of only 0.06~arcsec$^{-1}$ is also plausible given the dense environment that the lensing galaxies G1 and G2 reside and the gradient seems to increase in the direction of the peak in the X-ray emission. Again, weak lensing observations of the system would show if there is an external convergence gradient across the system, and whether this convergence gradient maps out the large-scale structure of the galaxy group associated with lensing galaxies G1 and G2. It would also be interesting to carry out resolved spectroscopy of the system (e.g. \citealt{czoske08}) to confirm the redshift of G2 and the lensed quasar, and to determine the velocity distribution of G1 to further constrain the lens models (e.g. \citealt{barnabe09}).

\section{Conclusions}
\label{conc}

We have presented new optical imaging and spectroscopy of the gravitational lens system B2108+213. Our observations have shown that the main lensing galaxy is the central galaxy of a massive group or poor cluster. Using both gravitational lensing and stellar kinematical analyses, we have carried out extensive modelling of the system and have come to the following conclusions.

i) By targeting a gravitational lens system with a wide image separation, we have successfully identified a large group of galaxies at a moderate redshift. Recent studies of the environments of gravitational lenses have found many are associated with groups of galaxies. However, these systems tend to have lensed image separations of order $\sim$\,1--1.5~arcsec, and are therefore, typically associated with poor groups, giving only limited information on the mass distribution of dark matter haloes in the mass regime between galaxies and massive clusters. Further investigations of group-scale dark matter haloes should target gravitational lens systems with image separations similar to or larger than $\sim$~4~arcsec.

ii) Our modelling of the inner part of the group mass distribution has found that the combined dark and baryonic matter density profile is much steeper (i.e. $\gamma' \ga$~2) when compared to what is predicted from dark-matter only simulations (i.e. $\gamma \sim$~1--1.5). This suggests that in the case of the group associated with B2108+213, the baryonic component within the inner 20~kpc of the halo centre dominates the mass distribution. B2108+213 may be a special case in this respect since there is a massive cD galaxy in the centre of the group. Therefore, it would be useful to find and study group systems without a massive central galaxy in the future.

iii) Our analysis of the gravitational lensing and stellar dynamics data finds that the inner matter density profile is on average isothermal within the Einstein radius, but is steeper (i.e. $\gamma' \sim$~2.5) at the position of the Einstein radius. A plausible explanation for this contradiction is the presence of an external convergence gradient across the lens system. The effect of this gradient is to change the flux-ratio of the lensed images and to bring the slope of the inner matter density profile determined from lensing to be consistent with what is predicted from the stellar dynamics of the lensing galaxy. The presence of a convergence gradient is certainly plausible considering there is a galaxy group associated with G1 and G2. Therefore, we find direct evidence for the environment of the lens system having an effect on the observed properties of the lensed images. Note that by not including a convergence gradient, the inner matter density profile is artificially steepened and this should be considered when modelling other group-scale systems in the future.

Our conclusions are based on the results from investigating only a single group system with gravitational lensing. Clearly, studies of many more group systems will be needed to test our findings for the B2108+213 group. Recently, there have been 13 group-scale gravitational lens systems identified from the Strong Lensing Legacy Survey (S2LS; \citealt{limousin09}). These groups span a large range of group masses and, apparently from maps of the luminosity distributions, a large scatter of group halo morphologies. It will be interesting to see whether this sample of strong lensing groups and poor clusters can be modelled without including an external convergence gradient.

\section*{ACKNOWLEDGMENTS}

The data presented herein were obtained at the W. M. Keck Observatory, which is operated as a scientific partnership among the California Institute of Technology, the University of California and the National Aeronautics and Space Administration. The Observatory was made possible by the generous financial support of the W. M. Keck Foundation. The results present herein were also based on observations collected with the NASA/ESA {\it HST}, obtained at STScI, which is operated by AURA, under NASA contract NAS5-26555. These observations are associated with program 9744. This work was supported by the European Community's Sixth Framework Marie Curie Research Training Network Programme, Contract No. MRTN-CT-2004-505183 `ANGLES'. This work was funded by the ANR grant ANR-06-BLAN-0067. LVEK, OC and SV are supported (in part) through an NWO-VIDI program subsidy (project number 639.042.505). CDF acknowledges support from {\it HST} program 9744.

\bsp

\label{lastpage}

\end{document}